\documentclass[fleqn,usenatbib]{mnras}

\usepackage[T1]{fontenc}
\usepackage{ae,aecompl}
\providecommand{\noopsort}[1]{}

\usepackage{graphicx}	
\usepackage{amsmath}	

\usepackage{amssymb}	
\usepackage[figure,figure*]{hypcap}
\usepackage{float}

\newcommand{\comm}[1]{}

\defcitealias{2019MNRAS.490.1539R}{R19}

\title[Analysis of individual cirrus filaments]{Prospects for future studies using deep imaging: Analysis of individual Galactic cirrus filaments}

\author[A. A. Smirnov et al.]{
Anton~A.~Smirnov,$^{1,2}$\thanks{E-mail:zeleniikot@gmail.com}
Sergey~S.~Savchenko,$^{1,2,5}$
Denis~M.~Poliakov$^{1,2}$
\newauthor
Alexander A. Marchuk,$^{1,2}$
Aleksandr~V.~Mosenkov,$^{3,1}$
Vladimir~B.~Il'in,$^{1,2,4}$
\newauthor
George~A.~Gontcharov,$^{1}$
Javier~Rom{\'a}n,$^{6,7,8}$
Jonah Seguine,$^{3}$
\\
$^{1}$Central (Pulkovo) Astronomical Observatory, Russian Academy of Sciences, Pulkovskoye chaussee 65/1, St. Petersburg 196140, Russia\\
$^{2}$Saint Petersburg State University, Universitetskij pr. 28, St. Petersburg 198504, Russia\\
$^{3}$Department of Physics and Astronomy, N283 ESC, Brigham Young University, Provo, UT 84602, USA\\
$^{4}$Saint Petersburg University of Aerospace Instrumentation, Bol. Morskaya ul. 67A, St. Petersburg 190000, Russia\\
$^{5}$Special Astrophysical Observatory, Russian Academy of Sciences, 369167 Nizhnij Arkhyz, Russia\\
$^{6}$Kapteyn Astronomical Institute, University of Groningen, PO Box 800, 9700 AV Groningen, The Netherlands\\
$^{7}$Instituto de Astrof{\'i}sica de Canarias, c/ V{\'i}a L{\'a}ctea s/n, E-38205, La Laguna, Tenerife, Spain\\
$^{8}$Departamento de Astrof{\'i}sica, Universidad de La Laguna, E-38206, La Laguna, Tenerife, Spain
}

\date{Accepted XXX. Received YYY; in original form ZZZ}

\pubyear{2022}
\begin{document}
\label{firstpage}
\pagerange{\pageref{firstpage}--\pageref{lastpage}}
\maketitle

\begin{abstract}
The presence of Galactic cirrus is an obstacle for studying both faint objects in our Galaxy and low surface brightness extragalactic structures. With the aim of studying individual cirrus filaments in SDSS Stripe~82 data, we develop techniques based on machine learning and neural networks that allow one to isolate filaments from foreground and background sources in the entirety of Stripe~82 with a precision similar to that of the human expert. Our photometric study of individual filaments indicates that only those brighter than 26 mag arcsec$^{-2}$ in the SDSS $r$ band are likely to be identified in SDSS Stripe~82 data by their distinctive colours in the optical bands. We also show a significant impact of data processing (e.g. flat-fielding, masking of bright stars, and sky subtraction) on colour estimation. \textcolor{black}{Analysing the distribution of filaments' colours with the help of mock simulations, we conclude that most filaments have colours in the following ranges: 0.55 $\leq$ $g-r$ $\leq$0.73 and 0.01 $\leq$ $r- i$ $\leq$ 0.33}. Our work provides a useful framework for an analysis of all types of low surface brightness features (cirri, tidal tails, stellar streams, etc.) in existing and future deep optical surveys. For practical purposes, we provide the catalogue of dust filaments.
\end{abstract}

\section{Introduction}
 Cirrus clouds are dust clouds usually observed at high galactic latitudes~($b \gtrsim 20^\circ$). They have filamentary wispy appearance and visually resemble the cirrus clouds observed in the Earth's atmosphere. 
Cirri were identified and studied over a wide range of wavelengths: in the infrared~\citep{Low_etal1984,Kiss_etal2001,Kiss_etal2003, Martin_eal2010,2011A&A...536A..22P,2012A&A...543A.123P,2020MNRAS.492.5420S}, optical~\citep{1955Obs....75..129D,1960Obs....80..106D,1972VA.....14..163D,1976AJ.....81..954S,1979A&A....78..253M,1985A&A...145L...7D, Ienaka_etal2013,2016A&A...593A...4M,Roman_etal2020}, and ultraviolet~\citep{1995ApJ...443L..33H,Gillmon_Shull2006,Boissier_2015,Akshaya_etal2019}. The cirri manifested in the visual and infrared, as well as in emission in the molecular CO and H$_2$ lines, were found to spatially correlate~\citep{Weiland_etal1986,deVries_etal1987,Gillmon_Shull2006, Ienaka_etal2013,Roman_etal2020}. 
\par
Cirrus clouds are unique objects both from theoretical and practical standpoints. They usually appear as numerous filaments rather than a cloud of a particular shape. Various studies~\citep{Bazell_Desert1988,Falgarone_etal1991,Hetem_Lepine1993,Vogelaar_Wakker1994,Elmegreen_Falgarone1996,Sanchez_etal2005,Juvela_etal2018, Marchuk_etal2021} of cirrus geometric properties proved that these clouds have a fractal nature. The fractal appearance of molecular clouds is thought to be due to the various physical processes that structure them: turbulence~\citep{2001ApJ...553..227P,Kowal_etal2007,Federrath_etal2009, Konstandin_etal2016,Beattie_etal2019a,Beattie_etal2019b}, shock waves~\citep{Koyama_2000}, colliding flows~\citep{Vazquez_Semadeni_2007}, and other factors, like the instability of a self-gravitating sheet~\citep{Nagai_1998} or various instabilities in non-self-gravitating clumps, which arise because of the presence of magnetic fields~\citep{Hennebelle_2013}.

\par
Considering the internal parts of cirrus, the optical spectrum of the diffuse galactic light (DGL), 
measured over 92.000 sky spectra from the Sloan
Digital Sky Survey~(SDSS,~\citealt{York_2000}), is found to be consistent with the spectrum of the scattered light~\citep{2012ApJ...744..129B, Chellew_2022} produced by a dust model of~\cite{Zubko_2004}.~\cite{Ienaka_etal2013} showed that this model can underestimate the correlation between the diffuse galactic light and the emission at 100 $\mu m$ by up to a factor of two if one measures the spectral properties of individual clouds. 
\par
From a practical standpoint, studies of cirrus are important for the following reason. With the progress in observational power and processing methods, it was shown that translucent cirrus clouds and other filamentary dusty structures are rather common inhabitants of sky regions at both high and low Galactic latitudes~\citep{Barrena_2018,2020MNRAS.492.5420S, Roman_etal2020}. Thus, they can interfere with studies of various extragalactic sources~\citep{Cortese_2010,Sollima_2010,Rudick_2010,Davies_2010,Duc_2018, Barrena_2018}. This problem was thoroughly discussed in~\cite{Roman_etal2020} in their study of optical cirrus based on SDSS Stripe~82 deep images~\citep{Abazajian_etal2009, Fliri_Trukillo2016}.~\cite{Roman_etal2020} identified and analysed sixteen clouds in the optical $g,r,i,$ and $z$ bands. One of the most important results of their work was that the cirrus clouds differ from typical extragalactic sources in terms of the optical colours $g-r$ and $r-i$. The authors suggested the following criterion, which allows one to distinguish cirrus filaments from any extragalactic objects based on the corresponding colours of specific image pixels: 
\begin{equation}
    (r - i) < 0.43 \times (g - r) - 0.06.
\label{eq:Roman}    
\end{equation} 
Since criterion~(\ref{eq:Roman}) includes only the optical colours, it provides an opportunity to distinguish the cirrus by means of optical data alone. Because various data sets have different resolutions, this criterion can become a valuable tool to
identify the cirrus presence in deep optical images. It is even more important when there is no complementary infrared data available, which is most frequently used to identify the presence of cirrus.
\par
Considering the nature of the suggested criterion, we should emphasise two important facts. First, the cirrus colours that appear in the inequality, are not the colours of each and every pixel of a cloud. Rather, they are the colours obtained from the linear fitting of the distribution of fluxes in the $(g, r)$ and $(r,i)$ planes (or by Gaussian plus Lorentzian fitting of the actual colour distributions) of a large sample of pixels. Such an approach implicitly assumes that a whole cloud, spanning several degrees of the sky, can be characterised by its unique colour, neglecting the possible variance of the colour over the different parts of the cloud. At the same time, we should note that almost every cirrus cloud consists of numerous filaments of different densities, surface brightnesses, etc. If the colour properties of the filaments vary too, it is important to verify the degree of their variance and the reliability of criterion~(\ref{eq:Roman}) as introduced by~\cite{Roman_etal2020} in this case.
\par 
The second important fact is that the spatial location of cirrus clouds were identified by~\cite{Roman_etal2020} by a visual inspection. In this work, we opt to take it a step further by using a more novel approach. Since cirrus clouds typically have similar wispy and filamentary structures, they are potentially good targets for automatic selection. For example, in a recent work by~\cite{2020MNRAS.492.5420S}, such structures were identified in Hi-Gal photometric survey data~\citep{Molinari_2010} based on their cylindrical-like shape, which is estimated using a Hessian matrix.  \textcolor{black}{A similar approach was used in~\cite{2016A&A...586A.135P} and~\cite{2022A&A...662A..96S} to study the relative orientation between the magnetic field and dust structures and between the HI filamentary structures and Galactic disc, respectively. In earlier works,~\cite{2013A&A...560A..63M} proposed to distinguish filaments (specifically those found in Galactic star-forming regions) using the decomposition of the images over a wide range of spatial scales. In~\cite{2015MNRAS.449.1782S}, authors applied a ridge detection technique and successfully extracted the filaments constituting a large ``integral shaped filament'' in Orion A North. \cite{2015MNRAS.452.3435K} suggested a complex approach, consisting of an $\arctan$ transformation of the image, Gaussian smoothing, and adaptive thresholding.} 
\par
In the present work, we adopt machine learning methods are suitable for an automatic search of cirrus clouds. Our goal is to test whether or not machine learning methods are suitable for automatic search of cirrus clouds. By identifying more cirrus clouds, we hope to acquire more reliable statistics of the cirrus photometric properties over different spatial scales. 
\par
The structure of the work is as follows. 
In Section~\ref{sec:data}, we describe the data and processing steps required for a measurement of cirrus colours: masking, the removal of the instrumental scattered light, and the cirrus filaments identification based on their visual appearance and the correlation with infrared data. 
In Section~\ref{sec:cirrus_nn}, we further improve the cirrus filaments identification with the aid of machine learning methods. Here we give the details about the setup of the method and the training of our neural network, and compare the results of the neural network and human identification. 
In Section~\ref{sec:sample}, we analyse the general properties of the sample of identified filaments.
In Section~\ref{sec:colour_measurement}, we discuss various pitfalls of the photometric analysis of the individual filaments and compare different approaches to the colour measurement using mock simulations. Here we also study how reliable colours are measured depending on the area and average surface brightness of the filament. 
In Section~\ref{sec:res}, we present the results of our colour measurement for a subsample of identified filaments and briefly discuss the spatial dependence of the colours on the galactic coordinates. 
We summarise our results in Section~\ref{sec:conclusions}.

\section{Data}
\label{sec:data}
We use the same Stripe 82 deep images as~\cite{Roman_etal2020}, where a large number of cirrus filaments/clouds can be distinguished simply by eye. The Stripe~82 data~\citep{Abazajian_etal2009} consists of 1100 fields covering a thin strip of the sky, 110 degrees wide ($-50^\circ < \alpha < 60^\circ$) and only 2.5 degrees in height ($-1.25^\circ<\delta<1.25^\circ$). The original raw fields were obtained using the 2.5-meter Apache Point Observatory telescope with an exposure time of one hour and a pixel scale of 0.396 arcsec. The fields were further stacked by~\cite{Fliri_Trukillo2016} and carefully processed in~\cite{Roman_Trujillo2018}, where the residues of the co-adding process were removed and improved sky-rectified images were obtained. \textcolor{black}{The resulting fields are two magnitudes deeper than the regular SDSS data.} The data from mentioned works is publicly available at \url{http://research.iac.es/proyecto/stripe82/}. Below we describe how we further processed the data from~\cite{Roman_Trujillo2018} to identify the cirrus filaments.

\subsection{Masking}
\label{sec:masking}
Images in the Stripe 82 survey contain numerous objects such as bright stars or galaxies, which have to be masked out before one can proceed with an analysis of Galactic cirri. In \cite{Roman_etal2020}, segmentation maps were created by running the {\small SEXTRACTOR} package~\citep{1996A&AS..117..393B} with various parameters to make initial mask images, which were further edited manually to include image artefacts.

To reduce our workload, we decided to use the mask images created by \cite{Roman_etal2020} for a set of Stripe~82 fields to train a neural network to generate masks for all Stripe~82  fields. We use an image-to-image algorithm based on the conditional adversarial network described in \cite{2016arXiv161107004I} as the neural network architecture. In this approach, two networks, a generator and a discriminator, are trained simultaneously. In the setting of our problem, the purpose of the generator is to
create a synthetic mask image based on a science image, and the goal of the discriminator is to determine if a particular mask image was created by a generator or by \cite{Roman_etal2020} (the discriminator  also has access to the optical images). During the training process, the generator learns to make more realistic masks to fool the discriminator. The discriminator in turn learns to more effectively distinguish between
real and synthetic maps to overcome the generator.

To create a training sample, we use optical images (in the $g$, $r$, $i$, and $z$ bands) and masks for these images provided by \cite{Roman_etal2020}, which were randomly cut into $256\times 256$ pixels segments (the input size of our networks). During the training process, we feed such cutout images to the generator and the discriminator and update their weights until the process converges. After that, the generative part of the network can be used to create masks for new (i.e. not covered by previous work) fields.

\begin{figure}
	\label{fig:masks_example}
    \begin{center}
        \includegraphics[width = 0.5\textwidth]{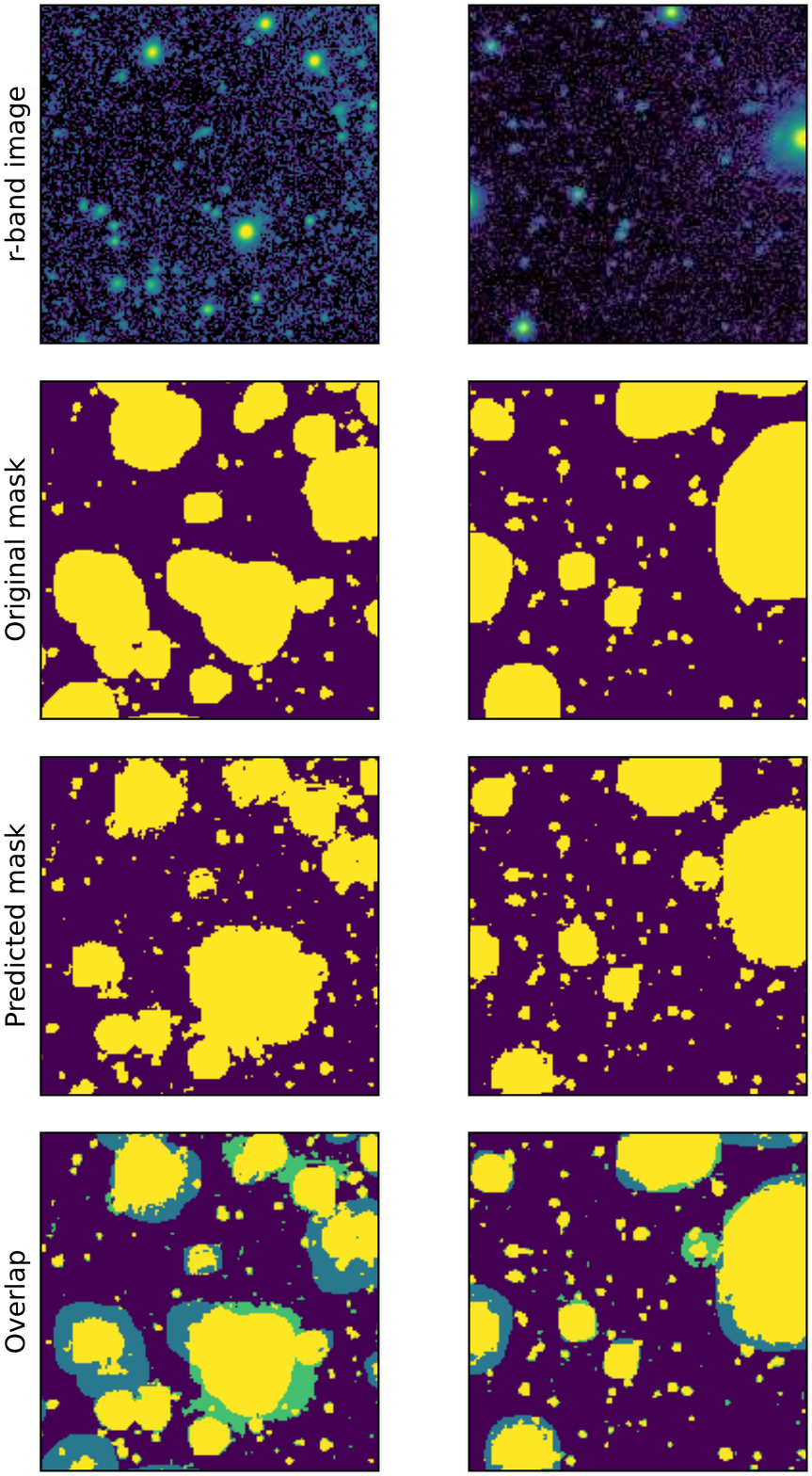}
    \end{center}
    \caption{
    Examples of the masks created by a neural network for two randomly selected patches of the Stripe~82. \textcolor{black}{Top panels: original images in the $r$-band, second row: masks made by \protect\cite{Roman_etal2020}, third row: masks generated by our neural networks. The bottom row shows the comparison of original and predicted masks: blue colour -- original masks, green -- predicted, yellow -- overlapping area of two masks.}}
\end{figure}

The results of the network training applied to a Stripe~82 field are shown in Fig.~\ref{fig:masks_example}: we show an $r$-band image, an original (the so-called ground truth) mask, and the prediction of the network for two random cutouts. It can be seen that, while the fine details of the generated masks differ, they generally cover all objects that present in the image. To  measure the similarity between the predicted and true mask, we use the intersection over union (IoU) metric:
\begin{equation} \label{eq:IoU}
	\mbox{IoU~} = \frac{\mbox{TP}}{\mbox{TP} + \mbox{FP} + \mbox{FN}}\,,
\end{equation}
where TP is the number of true positive pixel outcomes where the model correctly predicts the positive class, FP is the number of false positive pixel outcomes where the model incorrectly predicts the positive class, FN is the number of false negative pixel outcomes, where the model incorrectly predicts the negative class. For the trained network, the IoU median value for all the fields of the test sample is 0.69.

It should be noted that the network only deals with targets which are visible in the image. It is not aware of objects that may be outside of the image (but whose scattered light is present in the image), so the fine structure of the mask at the borders can be affected by this lack of data. For example, the faint wings of a bright star can be barely distinguishable in the image, but they would be covered by the mask if the centre of the star was visible. If the star is outside of the image provided to the network, the network is not aware of it and can miss the faint wings of the star. To deal with this problem, we only use central regions of the generated mask, and consider the data outside of this region as the context. To cover the whole field, we slide such a window
across it until the full mask for the field is created.

\subsection{Cirrus segmentation}
\label{sec:cirrus_picking_man}
A crucial moment in the cirrus analysis is detecting and selecting their locations in these images, i.e. selecting image pixels that are dominated by the cirrus scattered light and which do not contain other objects.
To do this, we applied the masks from Sect.~\ref{sec:masking} to the images to cover all non-cirrus objects and used a threshold of $29$ mag arcsec$^{-2}$ in the $r$ band (determined as the average $3\sigma$ limit for all Stripe~82 fields) to create a segmentation map of faint extended objects. Such segments constitute joint areas with
the surface brightness above the given limit. \textcolor{black}{Hereinafter, we define filaments as such joint areas. Thus, the filaments we identify here can be considered as separate segments of large cirrus clouds commonly studied in the literature.}

It turned out that even after applying the masks to the background objects, some other extended objects (not only cirrus) appeared in the image above the specified flux level. Among them are the faint extended wings of bright oversaturated stars and the reflections of such stars in the telescope optics, which manifest themselves as faint extended regions and can not be easily distinguished from cirrus by some easy-to-estimate parameter.

\begin{figure*}
	\label{fig:pipeline}
    \begin{center}
        \includegraphics[width = \textwidth]{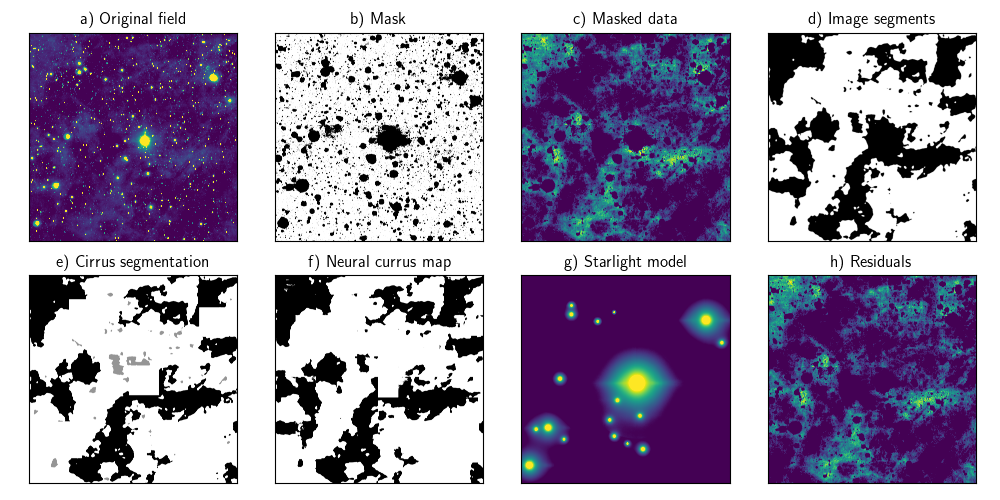}
    \end{center}
    \caption{Different stages of the cirrus segmentation applied to one Stripe~82 field. Panels are: a) the
        original field image in the $r$-band; b) mask of background and foreground objects generated using our neural network; c) masked original image with enhanced low surface
        brightness structures; d) image segments brighter than 29-th isophote in the $r$-band;
        e) cirrus/artefacts segmentation results; f) neural network generated cirrus mask;
        g) model of the scattered stellar light for bright stars; h) original masked image after
        the stars' model subtraction. See text for detailed information on the whole pipeline.}
\end{figure*}

To solve this problem, we decided to manually check every field by eye and individually select all of the regions that contained cirrus. We separated this list from that only contained the instrumental scattered stellar light.
Fig.~\ref{fig:pipeline} shows the stages of the manual cirrus segmentation for one field. The field itself is shown in panel a), while the field regions that are brighter than the 29-th isophote
on the $r$-band are shown in panel d) (this regions were computed using the
masked version of the image, so do not cover all visible stars). Panel e) of the figure shows the same segments separated into the ones
that cover cirrus regions (black) and the ones that cover image
artefacts (grey). We also removed the regions that are close
to the brightest stars (one in the middle of the panel e) to exclude them from consideration. To help with the selection, we also compare regions with infrared IRIS counterpart~\citep{IRIS}, available with lower resolution. In total we marked about 6.4 square degrees of the whole Stripe 82 area as cirrus (which is about 2\% of the survey area).

\subsection{Removing the scattered stellar light}
As noted by \cite{Roman_etal2020}, the images of Stripe 82 are contaminated by the
light of the extended wings of the brightest stars. The point spread functions (PSFs) have different
widths in different passbands (redder passbands have wider faint wings in their PSF), and also the
colours of stars are different. The result of these two factors is that different regions
of Stripe 82 fields have different background colours depending on the distance to the
bright stars, which significantly affect the measured cirrus properties.

To eliminate this problem, we follow the approach of \cite{Roman_etal2020} and fit the extended PSF\footnote{\url{http://research.iac.es/proyecto/stripe82/pages/advanced-data-products/the-sdss-extended-psfs.php}} models into locations of the brightest stars to subtract them from the
images and therefore remove the background colour variations. In this work, we use
the {\small TRACTOR} software \citep{Lang_2016}\footnote{\url{http://thetractor.org/}} to fit multiple extended PSF images prepared by \cite{Infante2020} to the Stripe 82 fields. In each field, we select all stars brighter than 15-th magnitude in $g$-band, similar to~\cite{Roman_etal2020}, and fit them iteratively starting with the single brightest star
and adding the next brightest star to the model at each step (computationally, this approach proved stabler than fitting all the stars in one step). During the fitting, we
mask out the regions that were marked as cirrus to exclude the influence of the cirrus on the
fitting of the stars (otherwise the cirrus contamination would be included in the model of the
extended PSF wings and removed after the model subtraction). Fig.~\ref{fig:pipeline}
shows the result of the stellar light modelling and subtraction for a randomly selected region that contains
both cirrus and some bright stars. Panel g) shows the model of the stellar light. Panel h) demonstrates the same region with the model subtracted.

\textcolor{black}{We note that this crucial step in the cirrus analysis pipeline requires a good knowledge of the extended  PSF wings. This problem is a typical obstacle for works in which  
low surface brightness structures are analysed \citep{Sandin_2014, Trujillo_2016, Karabal_2017}, and the proper PSF image should be created before proceeding to the actual analysis of the data (for example, \citealt{Rich_2019, Poliakov_2021})}.


\section{Automatic cirrus segmentation}
\label{sec:cirrus_nn}
Manual annotation of cirrus is very time-consuming for human experts. Careful annotation of a single $0.5^\circ\times0.5^\circ$ field in a semi-automatic approach may take up to 10 minutes. To investigate if the process of cirrus annotation can be fully automated and if the results of manual annotation can be further improved, we trained several U-Net \citep{2015arXiv150504597R} based networks. \textcolor{black}{In general, the U-Net architecture consists of two symmetrical paths: an encoder to capture context and a decoder to get precise localisation. The encoder follows the typical architecture of a convolutional network with repeating convolution and max-pooling operations. Every step in the decoder consists of an upsampling of the feature map followed by a convolution. Thus, the decoder increases the resolution of the output. To get localisation, the features from the encoder are combined with the upsampled features from the decoder via skip connections.}

\textcolor{black}{Originally, U-Net was proposed for biomedical image segmentation. This type of network architecture is successfully applied to various scientific and applied tasks such as medical image analysis \citep{2017arXiv171205053I, Ching142760, 10.1117/12.2293000, 2018SPIE10581E..1BI, Andersson2019SeparationOW, Nazem20213DUA}, cell biology \citep{Kandel2020PhaseIW}, and satellite image analysis \citep{2017arXiv170606169I}. It is also used in astronomical applications such as denoising, enhancing astronomical images \citep{2021MNRAS.503.3204V}, and stellar spectrum normalization \citep{2022A&A...659A.199R}.} In this section, we consistently describe these neural network models, through datasets (Sect.~\ref{sec:nn_datasets}), network architecture (Sect.~\ref{sec:nn_architectures}), and training methods (Sect.~\ref{sec:training_methods}). In Section~\ref{sec:nn_results}, we conduct our model analysis and discuss the results.

\subsection{Dataset for neural network training} 
\label{sec:nn_datasets}
\textcolor{black}{In Section~\ref{sec:cirrus_picking_man}, we carried out a manual identification procedure for cirrus filaments. Here we further translate the segmentation data to train an appropriate neural network. It is done in the following manner.} All pixels in Stripe 82 fields were annotated into 3 categories, in which $90.4\,\%$ of all pixels were background, $2.0\,\%$ were cirrus, and other extended sources the remaining $7.6\,\%$. The annotation for each field is stored in the corresponding annotation mask file. A value of $0$ for a mask's pixel denotes background, $1$ denotes cirrus, and $2$ denotes other extended sources. As the field image has a large size ($4553 \times 4553$ pixels), we employed square windows with smaller sizes for our models. It allowed us to decrease the time, memory capacity, and volume of manually annotated data required for training of network models.

To obtain the training, validation, and testing sample, we randomly chose three separate groups of fields consisting of 200, 50, and 100 fields, respectively. Here we briefly provide the main training and validation data pre-processing steps.
\begin{enumerate}
    \item We calculated the common $99.9$th percentile values for each optical band ($g, r, i$) separately for all training and validation fields (250 fields). Then we performed corresponding clipping. This moderates the problem of brightest pixels which reduces the image contrast, and therefore it increases the training efficiency.
	\item Next, we applied a natural logarithm transformation followed by min-max normalization to $[0:255]$ range.
	\item Then, we randomly chose $n_{\mathrm{tr}}$ square windows ($w \times w$ pixels) for each field in the training group and $n_{\mathrm{val}}$ for each field in the validation group. If the size of the obtained windows was too large for a current model, we resized each window to the spatial shape of the model input tensor ($w_{\mathrm{in}}, w_{\mathrm{in}}$), using \texttt{cv2.resize} method with \texttt{cv2.INTER\_LINEAR} interpolation. \textcolor{black}{As all considered architectures takes a 3 channel image input, the input tensor shape is ($3, w_{\mathrm{in}}, w_{\mathrm{in}}$).}
	\item The corresponding annotation mask's windows were obtained from the annotation mask files and resized, using \texttt{cv2.INTER\_AREA} interpolation.
	\item Lastly, during the formation of the input tensor we were applying min-max normalization to the $[-1:1]$ range and augmenting the data by symmetry group of square\textcolor{black}{. This group consists of $\pi/2$ rotations, reflections and their compositions (8 elements). Therefore,} this procedure increased the number of windows by a factor of 8.
\end{enumerate}

\subsection{Network architecture}
\label{sec:nn_architectures}
To resolve the task of cirrus annotation, we created several models based on the encoder-decoder U-Net-like
architecture \citep{2015arXiv150504597R}. All our experiments are conducted in the \texttt{TensorFlow2.x} framework \citep{tensorflow2015-whitepaper}. 
The precise manner in which each of these models described in this section is used to solve the issue of cirrus annotation is publicly available\footnote{\url{https://bitbucket.org/PolyakovD/cirrus_segmentation/src/master/}}.

\begin{figure}
	\label{fig:general_nn_architecture}
    \begin{center}
        \includegraphics[width = 0.5\textwidth]{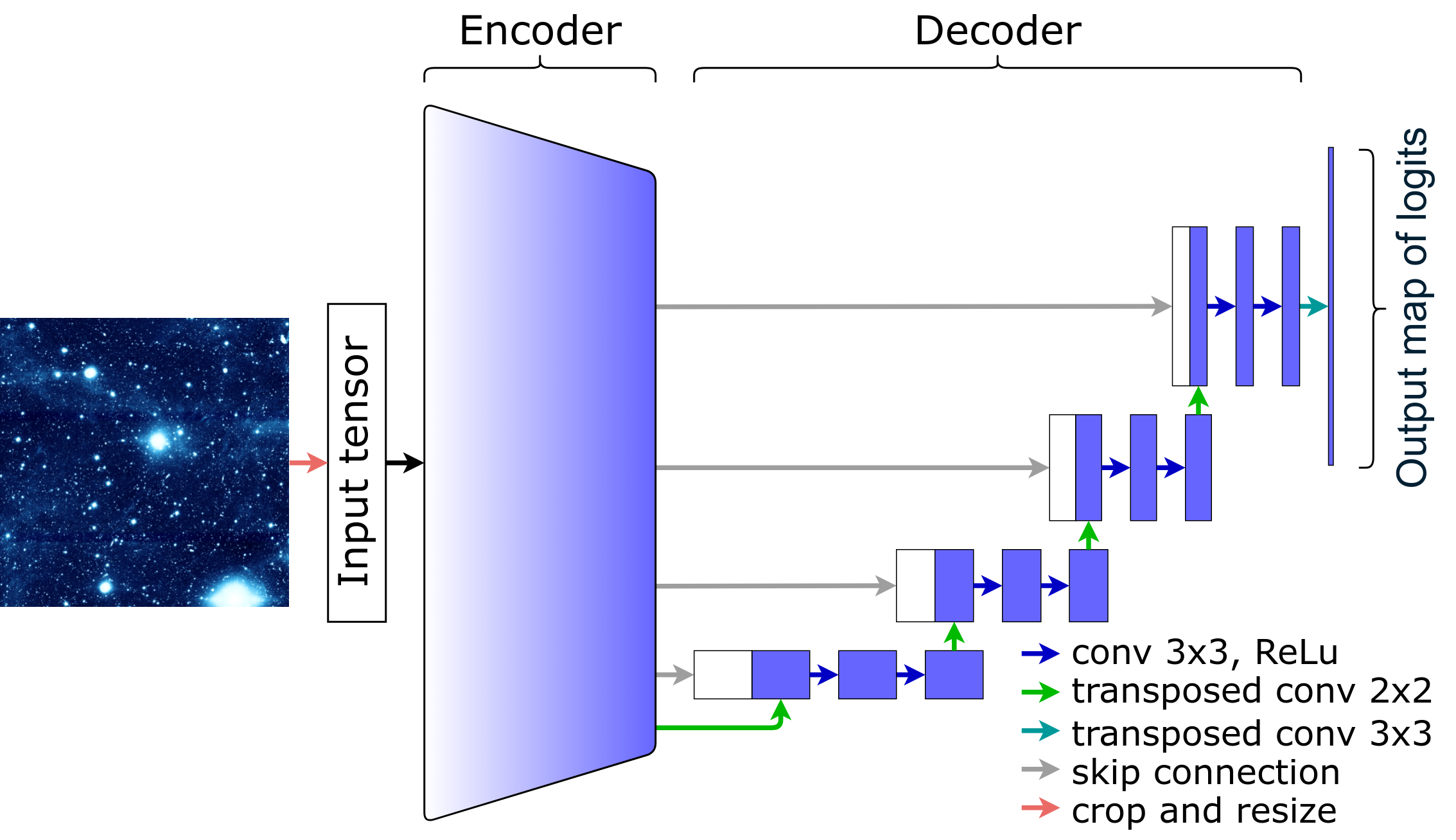}
    \end{center}
    \caption{The encoder-decoder architecture used in this work.}
\end{figure}

Fig.~\ref{fig:general_nn_architecture} shows a representation of the general architecture used. The key difference between \textcolor{black}{the} considered architectures is the encoder. As the encoder, we used ResNet50V2 \citep{2016arXiv160305027H}, MobileNetV2 \citep{2018arXiv180104381S}, and the classical U-Net encoder. 

The decoder architecture is identical for all models under consideration and consists of 4 steps (see Fig.~\ref{fig:general_nn_architecture}). Each of these steps have an upsampling of the feature map carried out with a $2\times2$ transposed convolution, a concatenation with the corresponding feature map from the encoder \textcolor{black}{(skip connection)}, and two $3\times3$ convolutions with zero padding, each followed by a ReLU. At the final layer, a $3\times3$ transposed convolution with stride~$= 1$ and zero padding is used to map each 64-component feature vector to the required number of classes.


\subsection{Training methods}
\label{sec:training_methods}
For each models under consideration, we used a sparse categorical cross-entropy loss function derived from \textcolor{black}{the} logits output tensor. To optimise the loss function, we employed the Adam optimization method with various learning rates $r$.

During \textcolor{black}{the} training experiments, we varied some parameters that influence the fitting process and the final model performance: the spatial shape of the input tensor ($w_{\mathrm{in}} \times w_{\mathrm{in}}$), the scale factor $s$ between the window size $w$ and the input tensor spatial size $w_{\mathrm{in}}$ ($w = s w_{\mathrm{in}}$), the training strategy (\guillemotleft training from scratch\guillemotright, \guillemotleft transfer learning\guillemotright, \guillemotleft fine-tuning\guillemotright), number of classes $n_{\mathrm{c}}$, class weights $\overline{\omega_{\mathrm{c}}}$, etc. We considered two cases for the number of classes, 3 classes which had been annotated in fields, and 2 classes when the \guillemotleft background\guillemotright\, class was extended by the \guillemotleft other extended sources\guillemotright\,class. In \guillemotleft transfer learning\guillemotright\,strategy, we took a pre-trained ImageNet dataset \citep{deng2009imagenet} encoder and froze it before the training process. In \guillemotleft fine-tuning\guillemotright\,strategy, we also used a pre-trained encoder but did not freeze it.

\textcolor{black}{To train our network models, we used a single NVIDIA GeForce RTX 3060 GPU. Batches consisted of 32 windows or 16 windows for models with the largest input tensor spatial size ($w_{\mathrm{in}} = 448$). We employed 30 epochs in all training experiments, since the lowest validation loss is reached in 10-20 epochs.}

\subsection{Experimental results and model analysis}
\label{sec:nn_results}
As demonstrated in Fig.~\ref{fig:pipeline} (panels e and f), the cirrus map generated by our best model is quite similar to the map obtained by human experts, and the model can successfully reproduce small cirrus filaments. 
To find this model, put the models through various comparative experiments. To compare models with each other, we use the IoU metric for the cirrus class (see eq.~\ref{eq:IoU}), which measures the similarity between the predicted and true cirrus. Human annotation performance yields a $0.67$ IoU for cirrus. This number was achieved by one expert on 100 random fields annotated by other experts of our team. Each of these fields was first annotated by one of the member of the group of experts. This annotation is considered as the ground truth annotation. The annotation of the single expert was then compared against this annotation. \textcolor{black}{The annotation procedure itself, carried out by a single expert, was done in a similar manner as it was done in Section~\ref{sec:cirrus_picking_man}, with the help of IRIS data~\citep{IRIS}.}

Quantitative results for different models and training methods are shown in Table~\ref{tab:nn_metrics}. We summarise the results of our experiments as follows.
\begin{enumerate}
	\item To find a more appropriate encoder, we conducted several experiments with various encoders. As one can see in Table~\ref{tab:nn_metrics}, models with the MobileNetV2 encoder demonstrate the highest performance (IoU $= 0.576$). Furthermore, these models are less resource-intensive and are more lightweight when compared to the others.
	\item The \guillemotleft fine-tuning\guillemotright\,strategy demonstrates the highest performance, but according to Table~\ref{tab:nn_metrics}, the advantage over models trained from scratch is insignificant.
	\item As one can see in Table~\ref{tab:nn_metrics}, models with moderately large windows ($w = 224, 448, 896$) are better than models with small windows. We assume that this might be related to the deficiency of semantic context in small windows relative to large ones.
	\item We also analyzed the models with 3 classes, but, as one can see in Table~\ref{tab:nn_metrics}, these models do not demonstrate an increase in performance when compared with the models with 2 classes.  
	\item The best of our models yields a 0.576 IoU. Since the advantage of human annotation is not great (0.67 IoU), it is possible to use this approach either the primary or supporting tool for annotating low surface brightness structures in deep astronomical images. It is remarkable that such an effective model was trained on only 250 fields out of 1100. The model makes a cirrus segmentation for one field in \textcolor{black}{about 25 seconds when running prediction on an AMD Ryzen 9 3900X 12-Core CPU and about 7 seconds when running on an NVIDIA GeForce RTX 3060 GPU.}

\end{enumerate}
\textcolor{black}{The fuzzy nature of cirri makes it difficult to translate the achieved IoU value into some transparent quality of the cirri detection. Even if some algorithm detects all the clouds in the image, the possible difference in the boundary threshold level will lead to an IoU value below unity. To give some perspective on the performance of our algorithm, we note that 89\% of the regions larger than 36 square arcseconds marked as cirri by humans have positive detections on the neural network inside their boundaries. Therefore, the vast majority of the cirri clouds can be detected by our network in an ``alert'' regime.}

\begin{figure*}
\begin{minipage}{0.9\textwidth}
\begin{center}
    \includegraphics[width=0.8\textwidth]{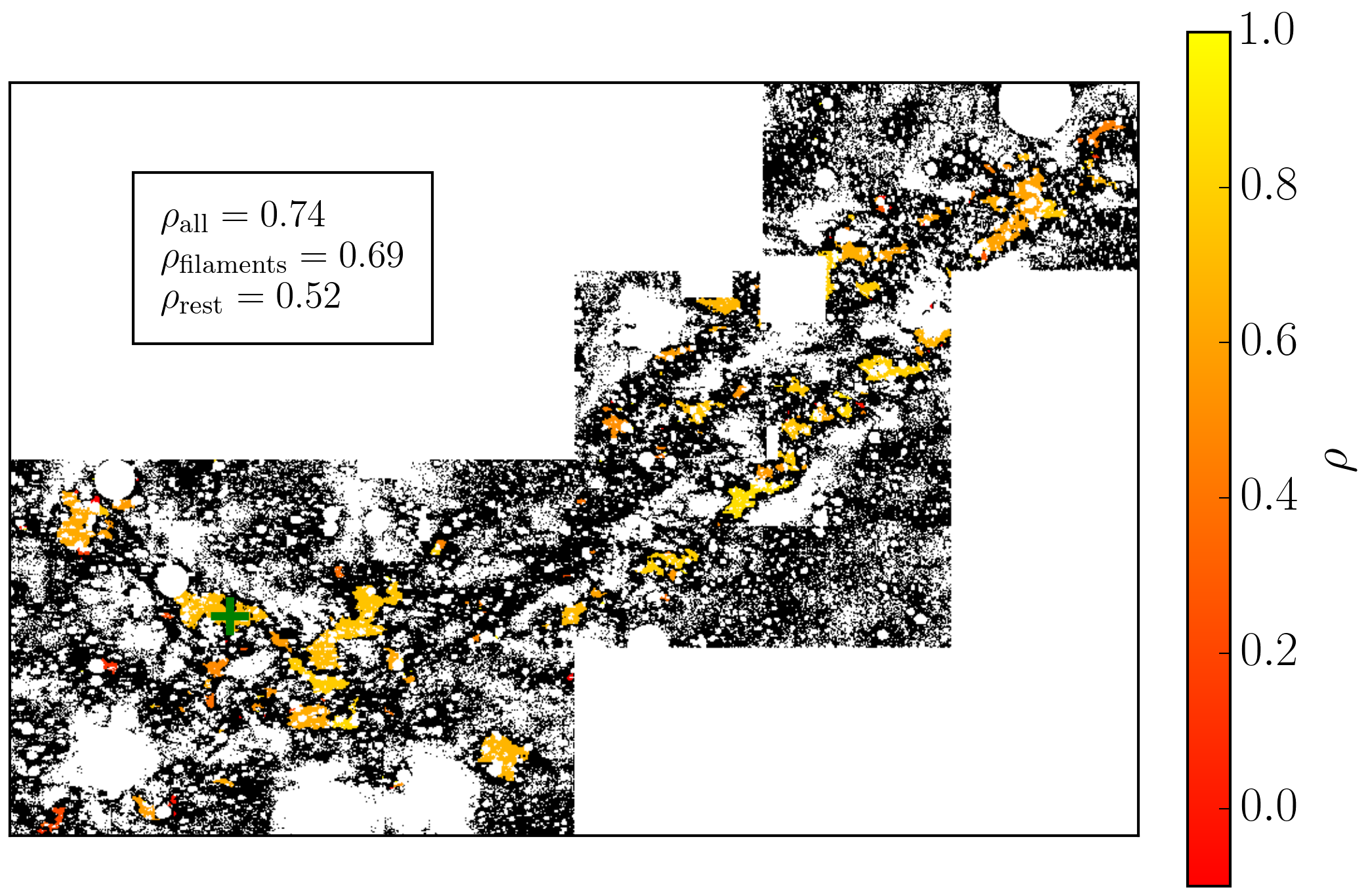}
\end{center}
\begin{center}
    \includegraphics[width=0.8\textwidth]{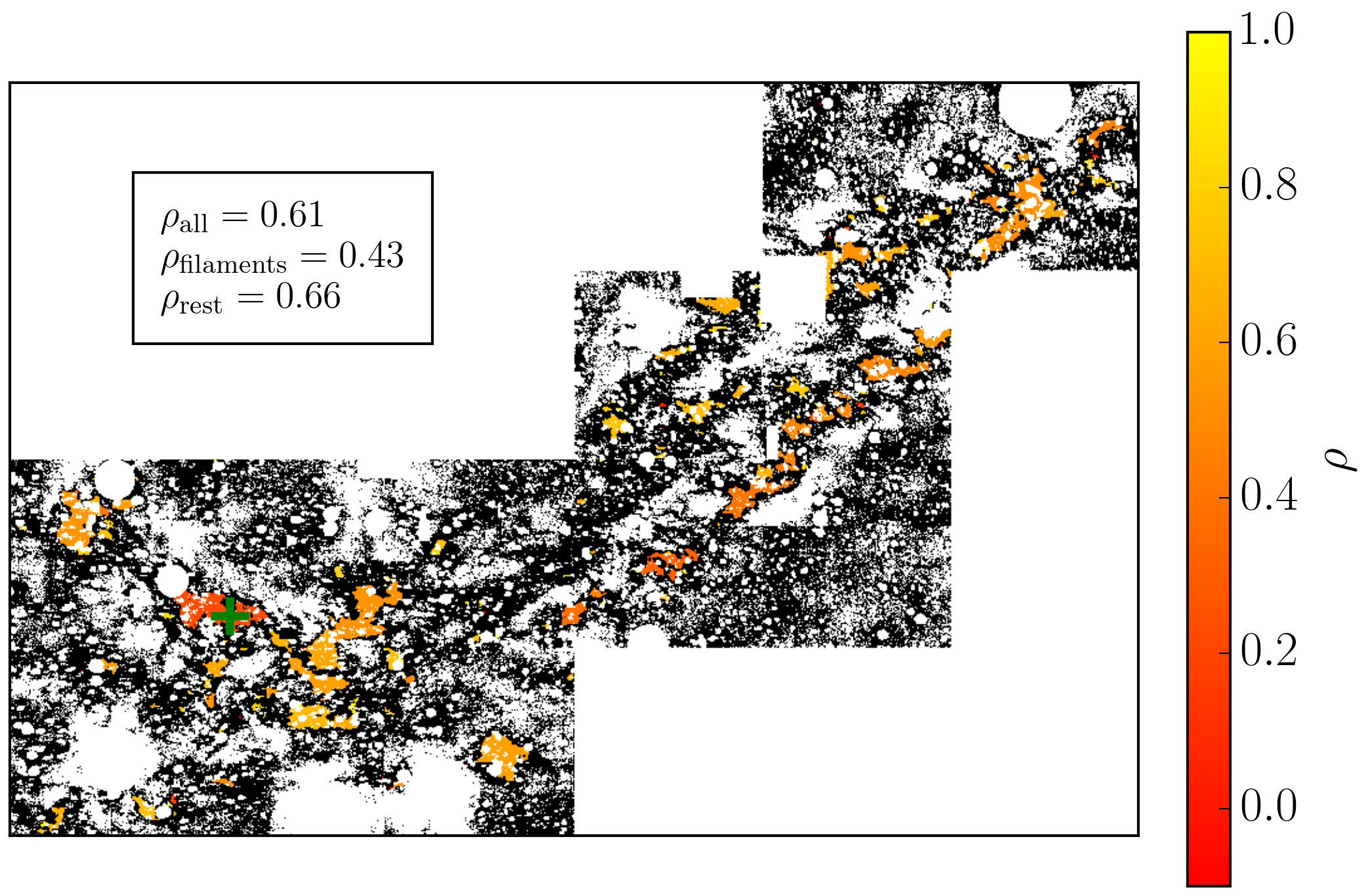}
\end{center}
\end{minipage}
\\
\caption{Colour coded correlation coefficients between Hershel and Stripe 82 $r$ band~(\textit{top} panel) and \textit{GALEX} and Stripe 82 $r$ band~(\textit{bottom} panel) for filaments of the cloud observed at $\alpha\approx 2.5^\circ$, $\delta\approx-0.25^\circ$. White areas correspond to the masked pixels. The green cross marks the filament that appears prominently in the IR data while not showing in the UV data.}
\label{fig:field5_cols}
\end{figure*}
\subsection{Correlation with IR and UV data}

The IRIS data which we use to support our identification of cirrus filaments in the optical have a low resolution of 90 arcsec. Therefore, it is instructive to verify how the fluxes are correlated between commonly used dust indicators, such as UV and IR, and optical for distinguished filaments if we consider more accurate data. For this purpose, we analyse only a single cirrus cloud, through one which is quite unique. It is located at $\alpha\approx 2.5^\circ$, $\delta\approx-0.25^\circ$ and appears in both the Hershel~\citep{Viero_etal2014} and \textit{GALEX}~\citep{Galex} datasets. The cloud is one of the richest cirrus clouds in Stripe 82 that was also studied by~\cite{Roman_etal2020} (their Field\#5). 
\par 
In Fig.~\ref{fig:field5_cols}, we present a map of individual filaments for this cloud. For each of the depicted filaments, we fill its area with the colour corresponding to the value of the correlation coefficient between \textit{Hershel} 250 $\mu m$ and Stripe~82 $r$-band data (\textit{top} panel) and between \textit{GALEX} far-ultraviolet (FUV) and the same $r$-band~(\textit{bottom} panel) data. \textcolor{black}{For each individual filament, a correlation coefficient $\rho$ is calculated by taking into account the fluxes in pixels within the area of the filament:
\begin{equation}
    \rho = \displaystyle\frac{\sum_{i=1}^n (x_i-\bar{x}) (y_i-\bar{y})}{\sqrt{\sum_{i=1}^n (x_i-\bar{x})^2} \sqrt{\sum_{i=1}^n (y_i-\bar{y})^2}},
\end{equation}
\textcolor{black}{where $x_i$ and $y_i$ are the fluxes in different bands, and $\bar{x}$ and $\bar{y}$ are their mean values, respectively. The summation is carried out over all pixels within the filament area.}
Thus, each filament is characterised by its individual correlation coefficient.} All analysed data are rebinned to a spatial scale of 12 arcsec to reduce the effect of the differences in their PSF, as well as possible pixel-scale spatial shifts of the datasets relative to each other. We also apply a very extensive mask, combining our mask produced by the neural network from Section~\ref{sec:masking}, the mask for this cloud from~\cite{Roman_etal2020}, and the mask obtained by cutting the bright sources in the UV and negative fluxes in the optical. Note that the corresponding correlation coefficients for each of the filaments are obtained using only the pixels within the area of the corresponding filaments. 
\par 
Fig.~\ref{fig:field5_cols} clearly shows that the dust emission in the IR and the scattered light in the optical are well-correlated ($\rho>0.5$) for most of the individual filaments, suggesting that we do indeed identify dust features. We also measured overall correlation coefficients using three separate sets of pixels: $\rho_\mathrm{all}=0.74$ (measured over all pixels in the depicted area), $\rho_\mathrm{filaments}=0.69$ (measured over the pixels within the filaments), and $\rho_\mathrm{rest}=0.52$ (measured over the pixels that are outside of the filament boundaries). The fact that the $\rho_\mathrm{rest}\gtrsim0.5$  and $\rho_\mathrm{all}>\rho_\mathrm{filaments}$ may indicate that we miss some part of the cirrus in the optical. This also clearly follows from the comparison with the \textit{GALEX} data~(\textit{bottom} panel of Fig.~\ref{fig:field5_cols}). There, $\rho_\mathrm{filaments}=0.43$ is smaller than both $\rho_\mathrm{all}$ and $\rho_\mathrm{rest}$. Although, as can be seen, for many filaments $\rho$, is still close to $0.5$. At the same time, for some filaments, there is no correlation with \textit{GALEX}, although such a correlation is present when using the Hershel data for the same filament (see a big filament in the lower left corner of both maps marked by a green cross). We should note that a qualitatively similar discrepancy regarding infrared and UV data was noted by~\cite{Boissier_2015} in their study of cirrus in the Virgo cluster. The authors found that some cirrus regions that appear in the FUV maps are not visible in the infrared or Planck maps and vice versa.
\par 
The presented comparison with other data sets shows that the areas which were distinguished as cirrus filaments by our neural network do indeed host dust features. The comparison also indicates that we do not identify a portion of the cirrus in the optical. It is hard to estimate exactly how much of the filaments we miss, but this is expected because we are limited by the depth of the data and, therefore, we cannot identify dim filaments which can appear more prominently in the IR and UV. 

\section{Resulting sample of filaments}
\label{sec:sample}

\begin{figure*}
    \centering
    \includegraphics[width=0.8\textwidth]{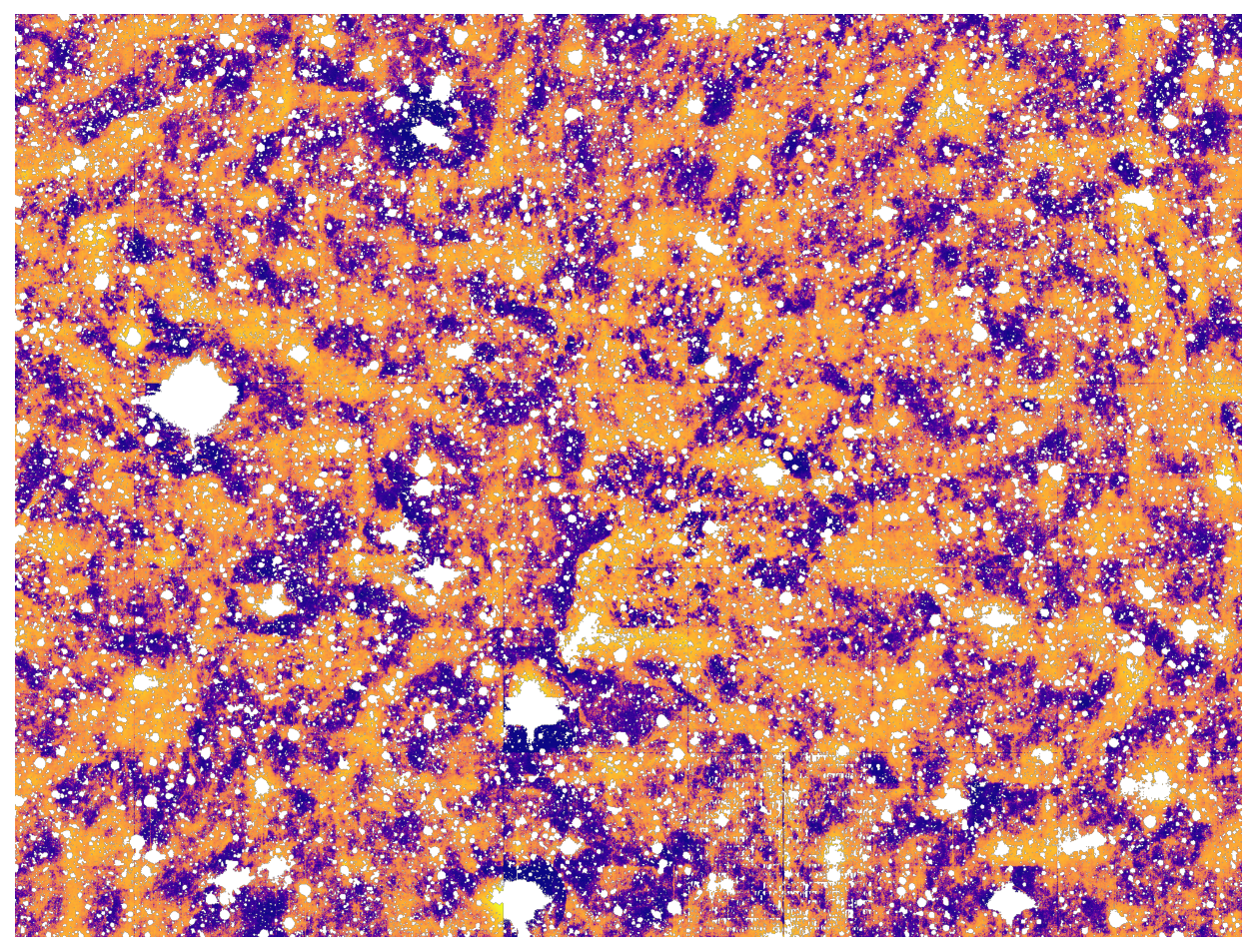}
    \includegraphics[width=0.8\textwidth]{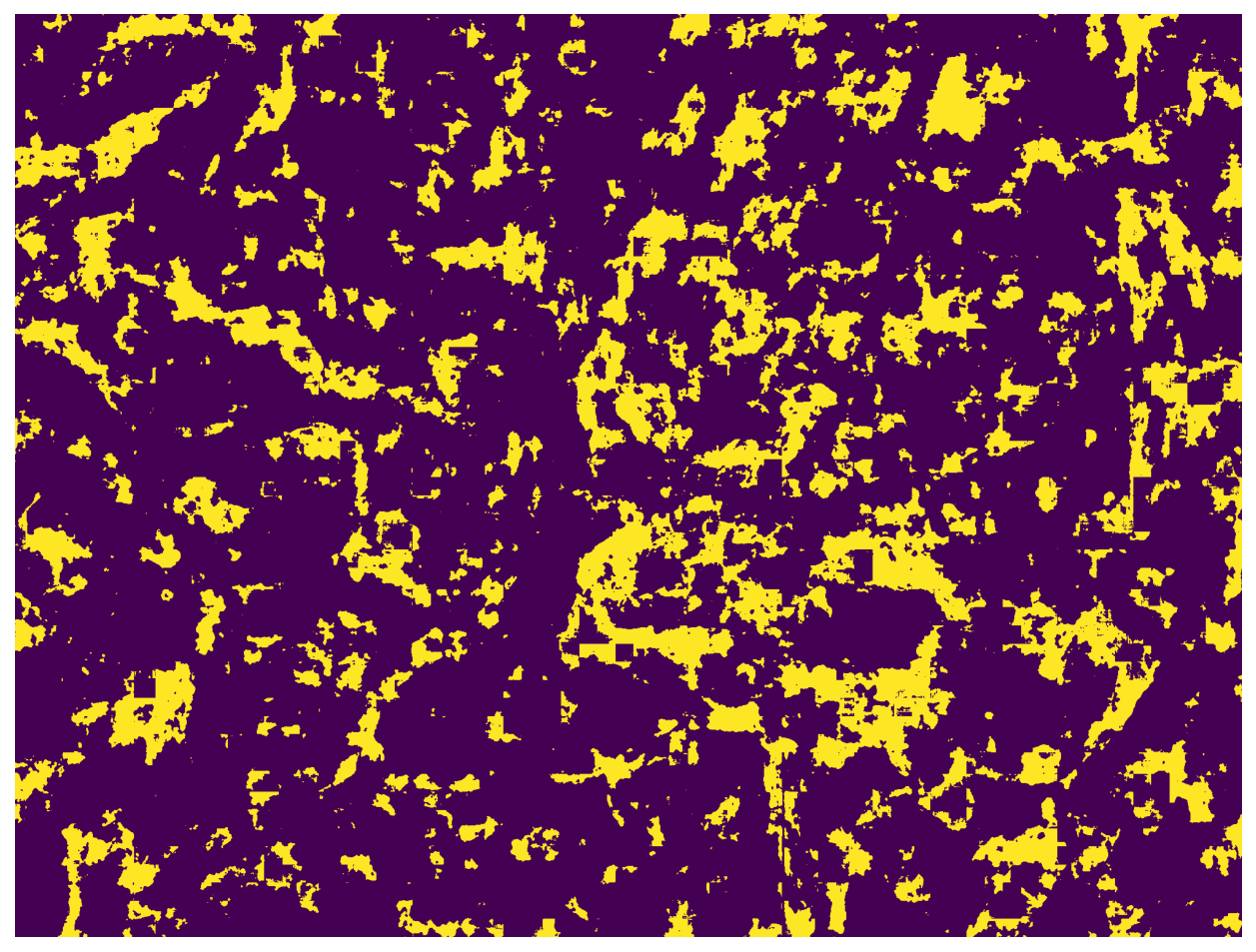}
    \caption{Cirrus rich area at $\alpha=55^\circ-60^\circ$ containing about one hundred of Stripe 82 fields ($\sim5$ square degrees): intensity map (\textit{top} panel) and cirrus map created by neural network (\textit{bottom} panel).}
    \label{fig:beatiful_cirrus}
\end{figure*}

The resulting sample of filaments identified by our neural network consists of about $5\cdot 10^5$ spatially separated regions. The total area covered is about 6.6 square degrees, which is greater than that obtained via manual picking by $0.2$ square degrees. \textcolor{black}{For illustrative purposes, we present Fig.~\ref{fig:beatiful_cirrus}, which shows a cirrus rich area at $\alpha\approx55^\circ-60^\circ$ (one of the ends of Stripe 82). The whole presented area contains about one hundred original Stripe 82 fields ($\sim5$ square degrees). The top panel of the figure shows an intensity map with the masking and source's subtracting carried out, while the bottom panel shows the areas identified by neural network as cirrus filaments.} In this section, we describe some general properties of the filaments' sample, as well as some preliminary steps that must be taken before analysing the colours of the filaments.
\par 
First of all, at the original pixel scale, the data is dominated by the noise that ever-present in astronomical images. To facilitate the analysis of dust colours, we reduced the noise contribution by rebinning each field's images to a spatial resolution of $6$ arcsec, similarly to~\cite{Roman_etal2020}. \textcolor{black}{They decided on that resolution in that work as a compromise between optimal spatial resolution and image depth. To make the comparison between the results in~\cite{Roman_etal2020} and this work clearer, we decided to use the same spatial resolution in present work.} At this step, we assume that if half of the small pixels with a scale of 0.396 arcsec (which constitute the large 6 arcsec pixel) are initially marked as being dominated by scattered cirrus light, the large pixel should also be marked as dominated by cirrus. In the other case, the large pixel is simply removed from the analysis. As a result of this procedure, the filaments' number is significantly reduced to 23290, while the total marked area does not change (the same 6.6 square degrees). The decrease in the number of filaments is explained by the fact that the original sample contains a significant amount of small features with a spatial scale of only a few pixels. When we rebin the images, such features are either removed from the analysis or merge into a single filament with a larger size. 
\par 
\begin{figure*}
\begin{center}
\begin{minipage}{0.8\textwidth}
\includegraphics[width=0.99\textwidth]{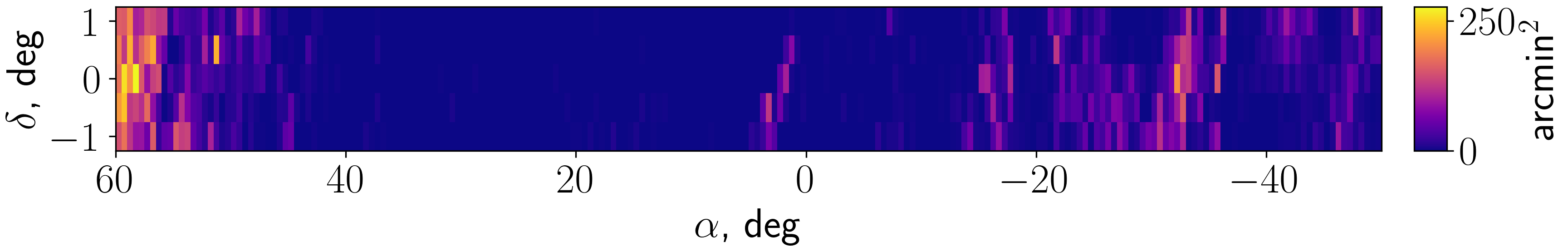}
\end{minipage}%
\end{center}
\caption{Distribution of identified filaments across the sky plane. Each small rectangle in this map corresponds to one of the original Stripe 82 fields \textcolor{black}{each with an area of about 900 arcmin$^2$}, while the colour of the rectangle corresponds to the total area within the field marked as dominated by cirrus.}
\label{fig:fil_spatial}
\end{figure*}

\begin{figure}
    \begin{center}
        \includegraphics[width = 0.5\textwidth]{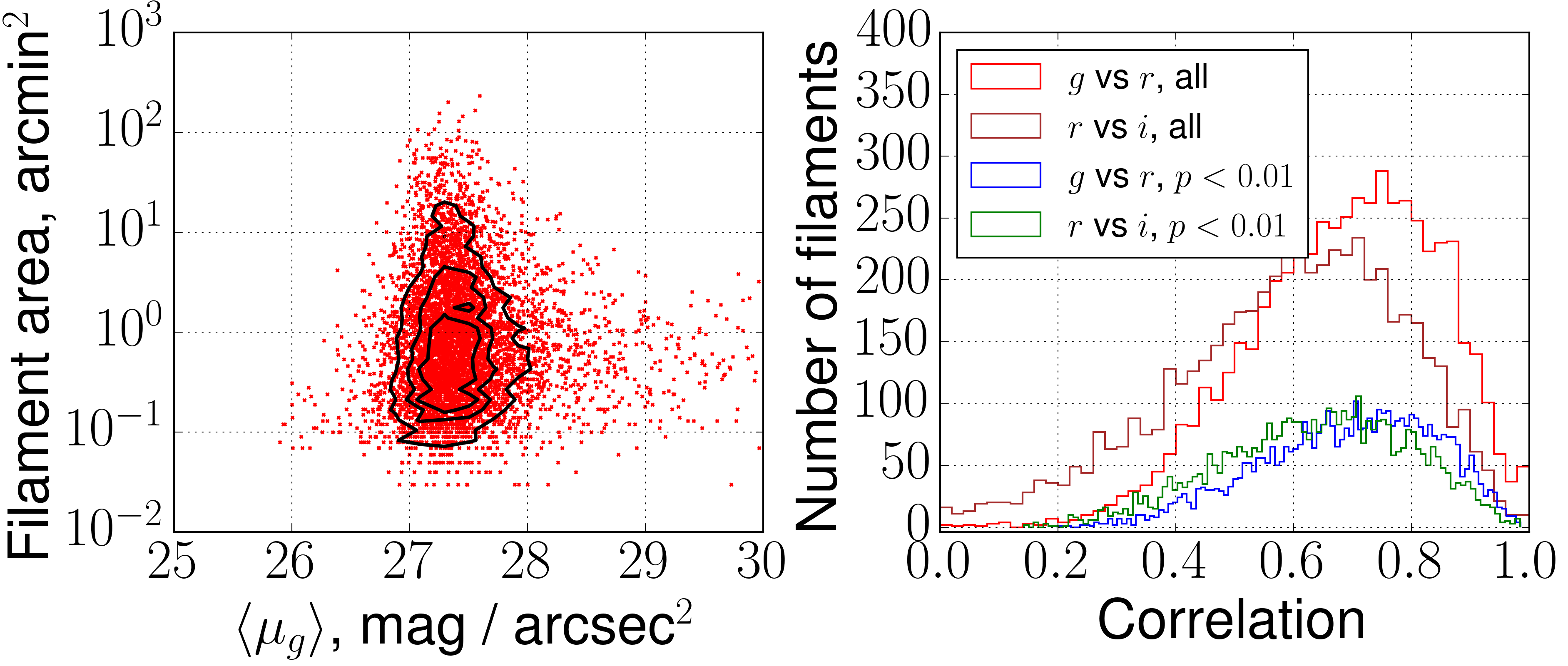}
    \end{center}
    \caption{Left panel: distribution of filaments over the average surface brightness in the $g$ band and the cloud area. Right panel: distributions of the correlation coefficients between the fluxes for all filaments (red and brown lines) and a subsample of filaments with $p$-value smaller than 0.01 for the measured correlation (blue and green lines).}
    \label{fig:correlations}
\end{figure}

Fig.~\ref{fig:fil_spatial} shows the spatial distribution of the filaments over Stripe 82 after the rebinning has been carried out, and Fig.~\ref{fig:correlations} presents a variety of statistics, such as the distribution of filaments over the average surface brightness and the area. In the right panel of the figure, we display the distribution of filaments by the correlation between the $(g,r)$ and $(r,i)$ pairs of the optical bands. We do not consider data in the $z$ band because it is less deep and our observations showed no correlation at all between the $z$ band and others in many small filaments. We also depict the distributions for a subsample of $4575$ filaments where the correlation is reliably measured, that is the subsample includes only those filaments with $p$-value smaller than 1\% (which means that a random data has less than a 1\% chance of resulting such a strong correlation). Below, we present the results of the colour measurements for filaments only from this subsample. Thus, the total number of analysed filaments is $4575$ and the total area is about 4.5 square degrees (70\% of the original area).  
\par 
The left panel of Fig.~\ref{fig:correlations} emphasises the difference between the current analysis and those executed previously, such as in~\cite{Guhathakurta1989} and \cite{Roman_etal2020}. In these works, authors considered distributions of the fluxes for all pixels within an area of several or more square degrees. The typical area of the filaments considered  in this work is smaller by an order of a magnitude. As for the surface brightness, the majority of our filaments are dim features with $\langle \mu_g \rangle> 27$ mag arcsec$^{-2}$. These differences imply that special care must be taken if one tries to measure the colours of such features. We thoroughly discuss this problem in the next section. 

\section{Colour measurement}
\label{sec:colour_measurement}


\begin{table*} 
	\caption{Details of the mock simulations used over course of the present work. The first column shows a simulation ID, columns to thorough five present the limits of the physical parameters of the squares representing the cirrus filaments in simulations, namely area, surface brightness, and colours. ``U'' (uniform) and ``RL'' (real-like) abbreviations, given in brackets, indicate whether the adopted distribution for each particular parameter is a uniform one (``U'') or specifically prepared to resemble the corresponding distribution for the real filaments (``RL'', see Fig.~\ref{fig:correlations}, \textit{left}). The sixth column gives the description of the background into which the squares are injected, while the seventh column gives a short description of the problem, which is solved using each particular simulation.} 
	\centering
    \begin{tabular}{l l c c c c c c c c}
        \hline \hline
        Name  & Area, arcsec$^2$ & $\mu$, mag/arcsec$^2$ & $g-r$ & $r-i$  & Background &Purpose  &  \\
          \hline
        S1   & 10$^2$-10$^5$ (U) & 25-29 (U) & 0.1-0.8 (U) & 0.1-0.8 (U)& Stripe 82 fields & comparison of colour measurement procedures \\
        S2   & 10$^2$-10$^5$ (U) & 25-29 (U) & 0.1-0.8 (U)& 0.1-0.8 (U)& Gaussian noise &  estimation of the background effects \\
        S3  & 10$^2$-10$^5$ (RL) & 26-30 (RL) & 0.5-0.7 (U)& 0.0-0.2 (U)&  Stripe 82 fields & estimation of colour spread for a ``point source'' \\
        S4  & 10$^2$-10$^5$ (RL) & 26-30 (RL) & 0.0-1.5 (U)& -0.5-1.0 (U)&  Stripe 82 fields & finding colours of the real filaments \\

    \end{tabular} 
    \label{tab:sim}
\end{table*}

\begin{figure*}
    \centering
    \begin{minipage}{0.22\textwidth}
    \includegraphics[width=0.99\textwidth]{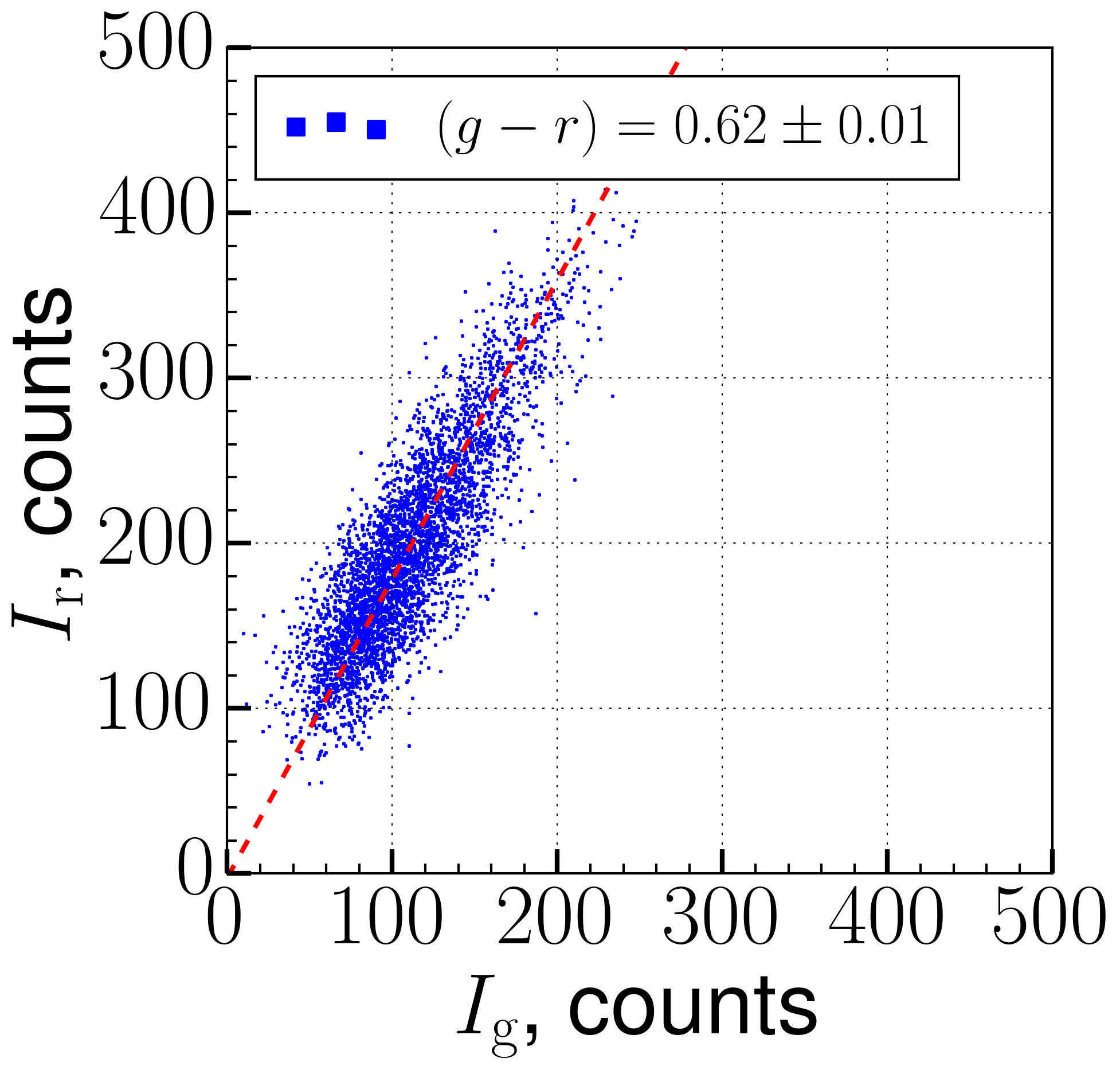}
    \end{minipage}%
    \begin{minipage}{0.22\textwidth}
    \includegraphics[width=0.99\textwidth]{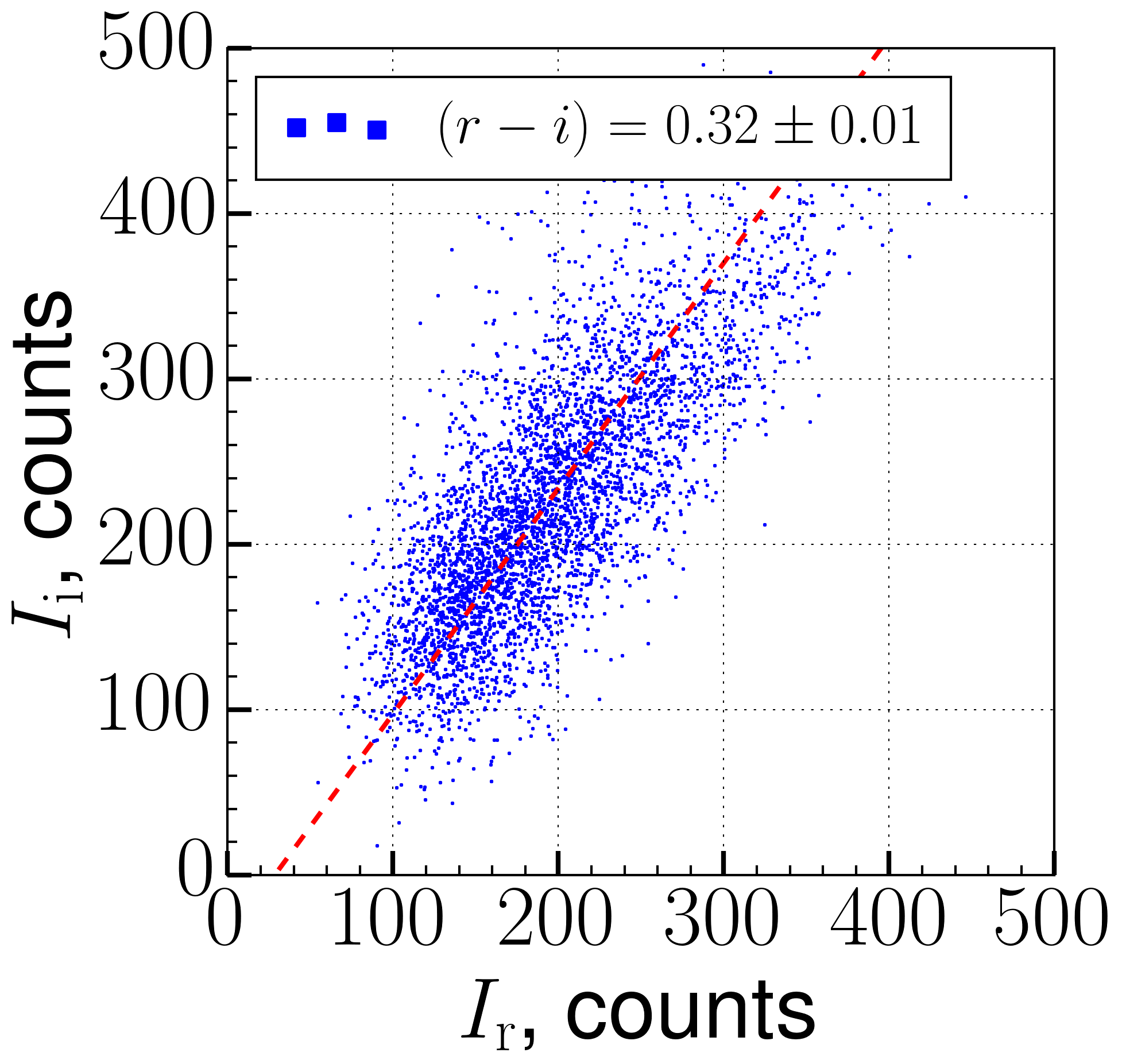}
    \end{minipage}%
    \begin{minipage}{0.27\textwidth}
    \includegraphics[width=0.99\textwidth]{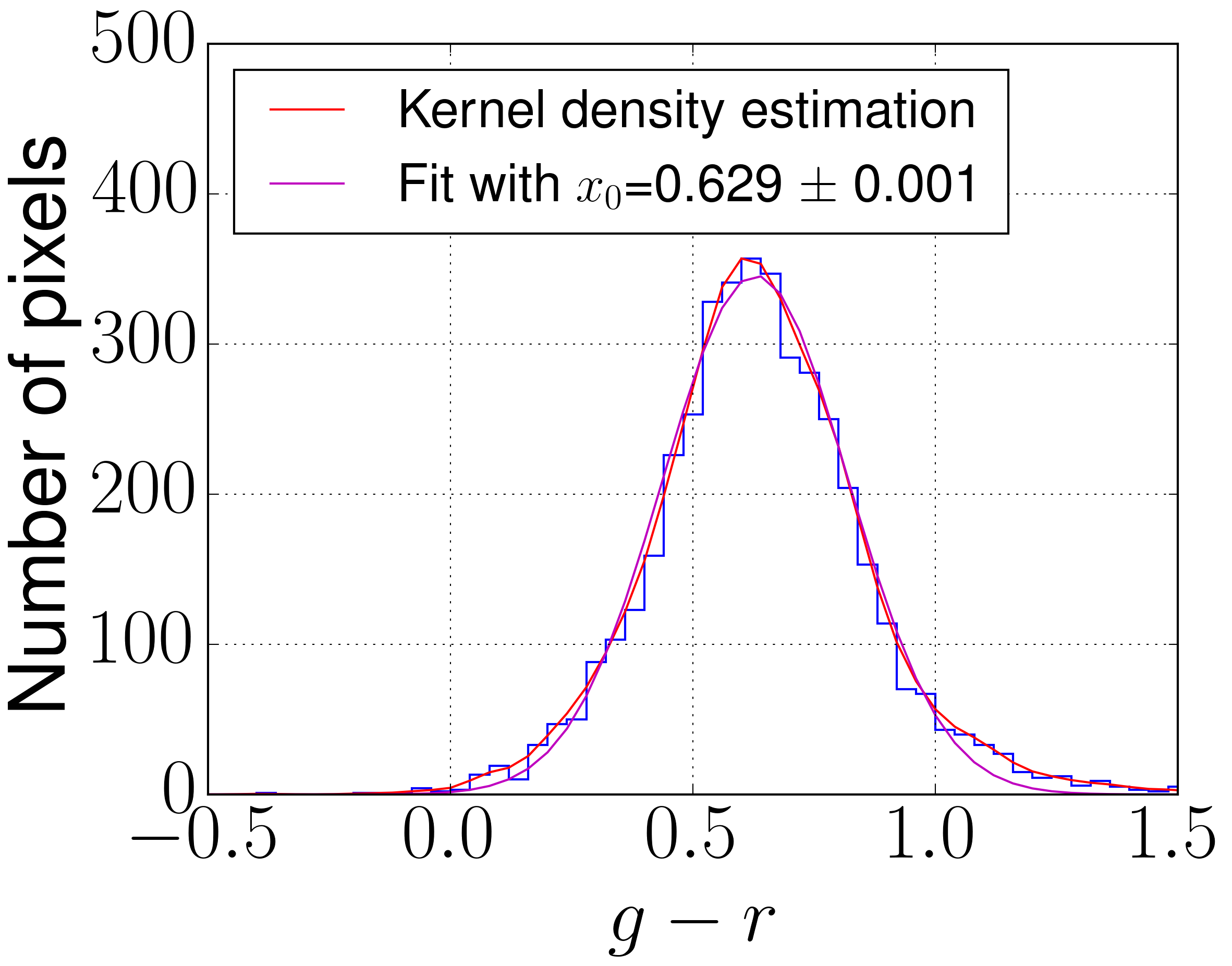}
    \end{minipage}%
    \begin{minipage}{0.27\textwidth}
    \includegraphics[width=0.99\textwidth]{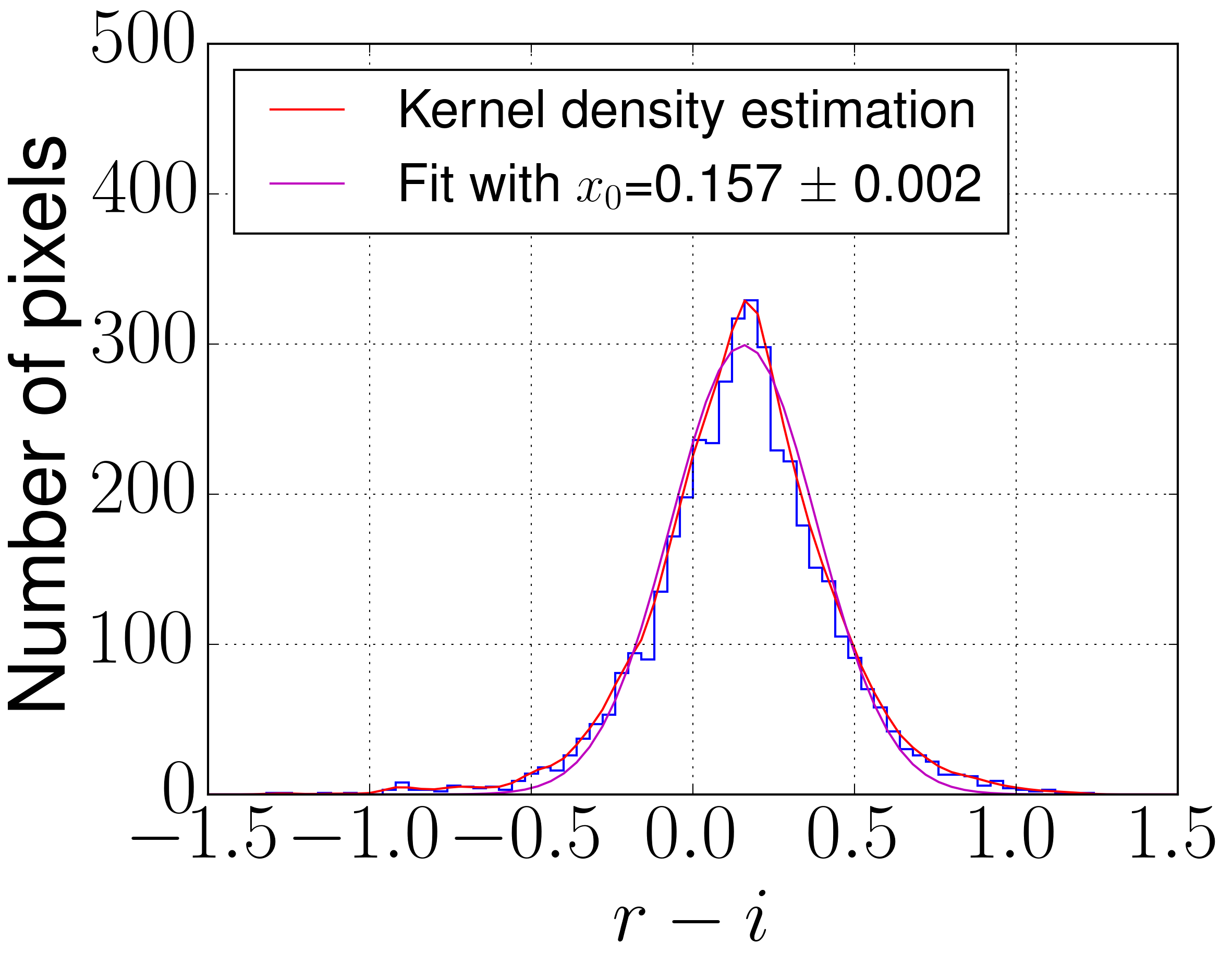}
    \end{minipage}
    \caption{Approaches used for measuring the colours of the filaments in the present work applied to the largest filament from Fig.~\ref{fig:field5_cols}. Two left panels: the linear fitting method. Two right panels: Gaussian fitting of the colour distributions of the filament pixels.}
    \label{fig:fil_measurement}
\end{figure*}

\begin{figure*}
\begin{minipage}{0.25\textwidth}
\includegraphics[width=0.99\textwidth]{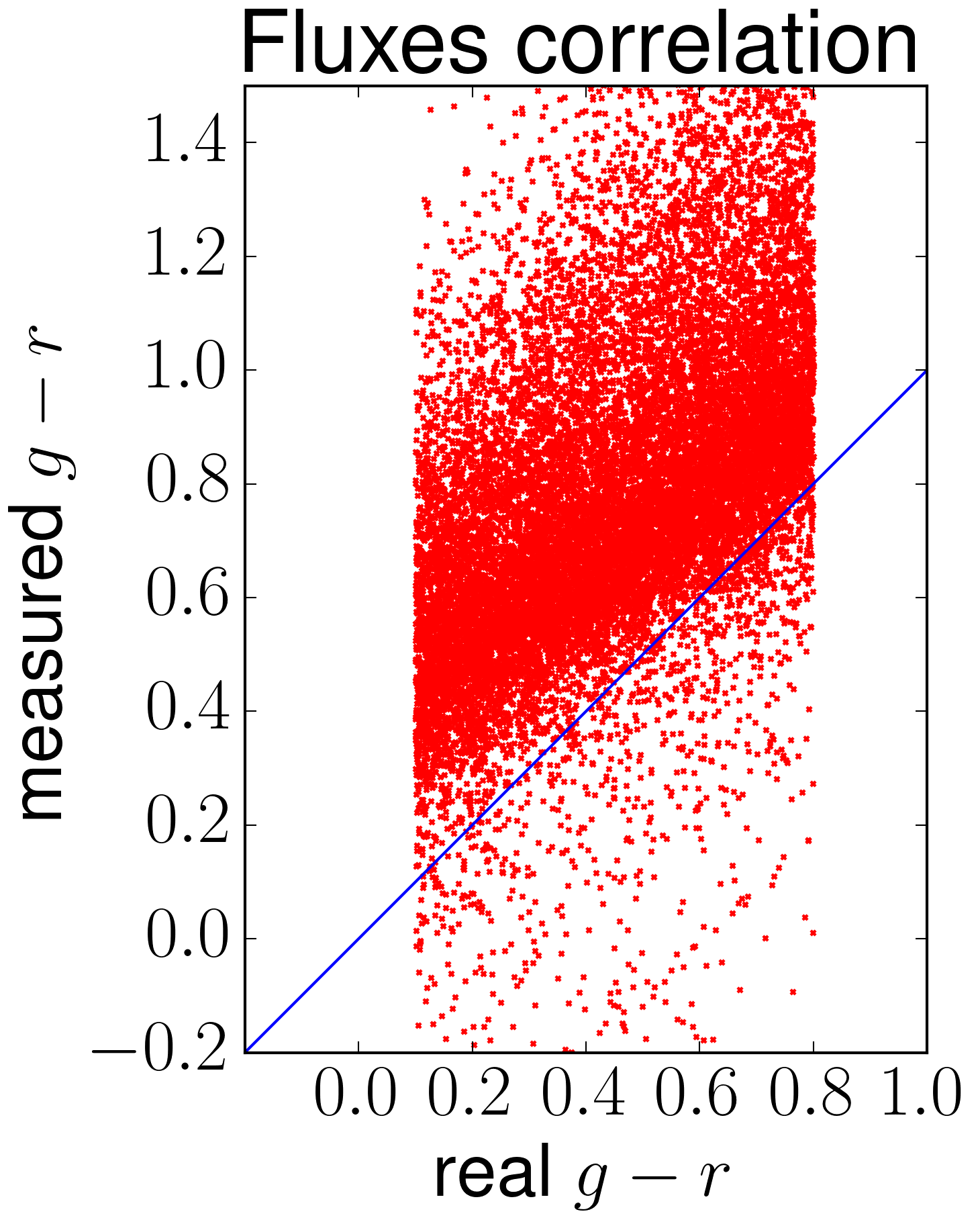}
\end{minipage}%
\begin{minipage}{0.25\textwidth}
\includegraphics[width=0.99\textwidth]{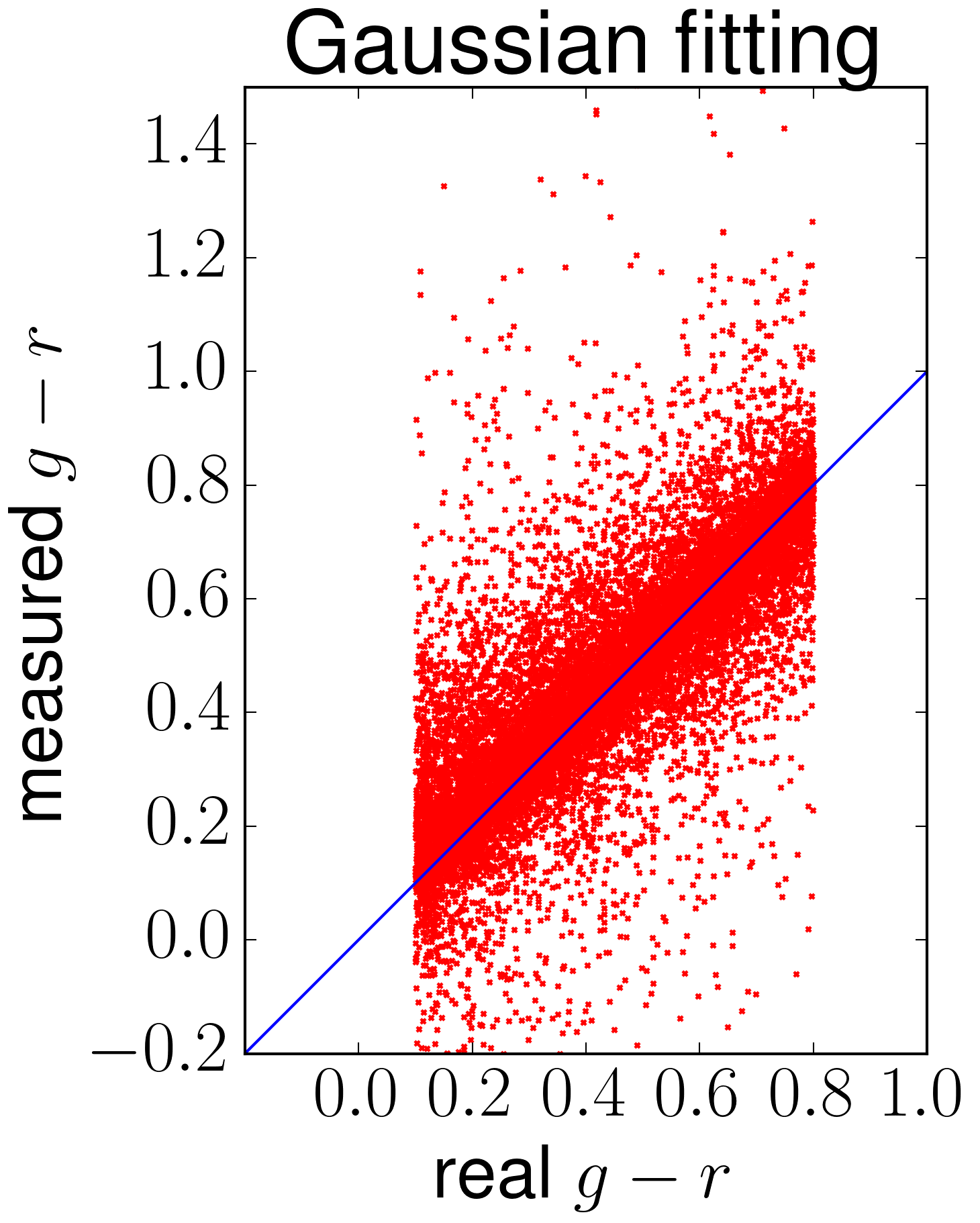}
\end{minipage}%
\begin{minipage}{0.25\textwidth}
\includegraphics[width=0.99\textwidth]{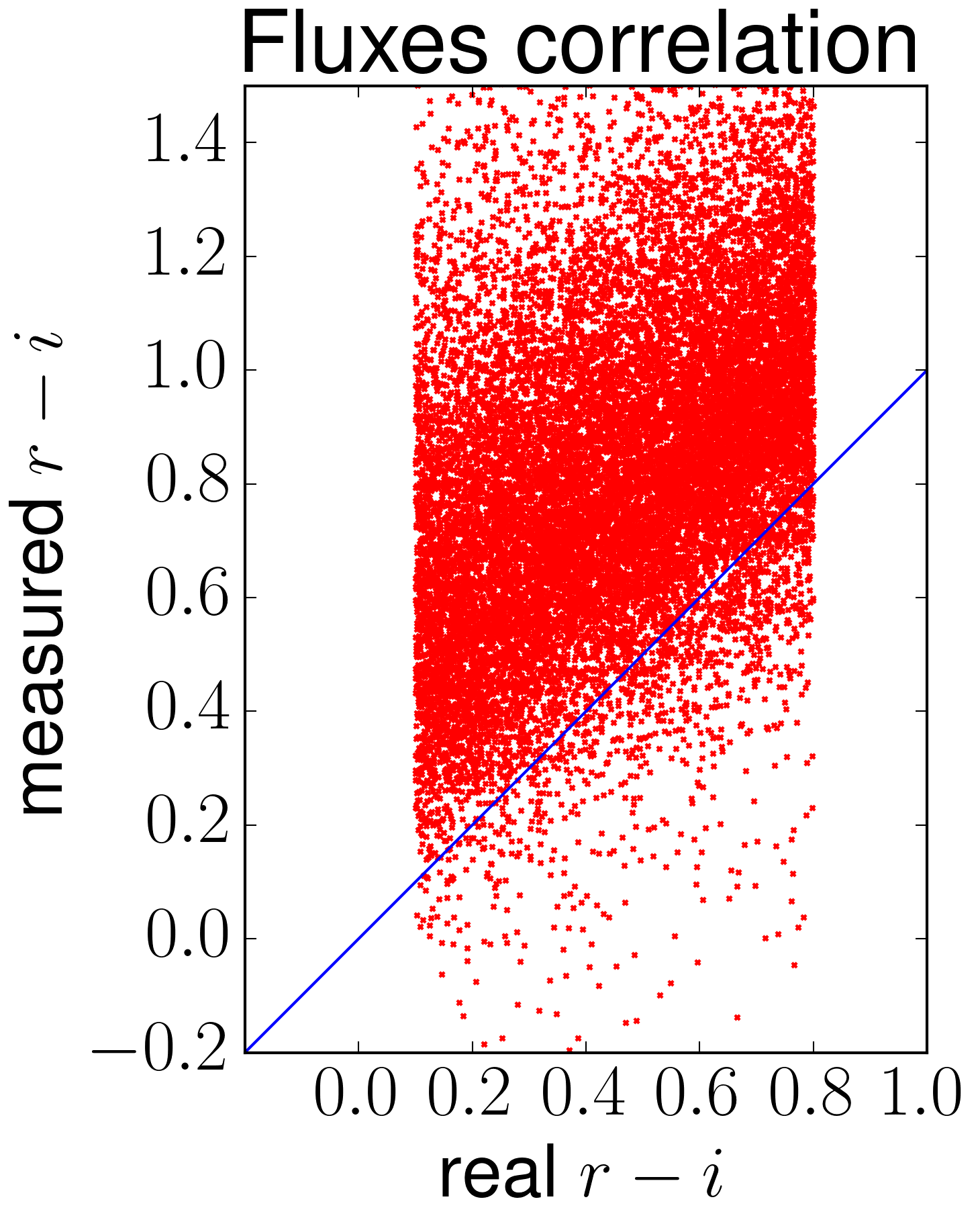}
\end{minipage}%
\begin{minipage}{0.25\textwidth}
\includegraphics[width=0.99\textwidth]{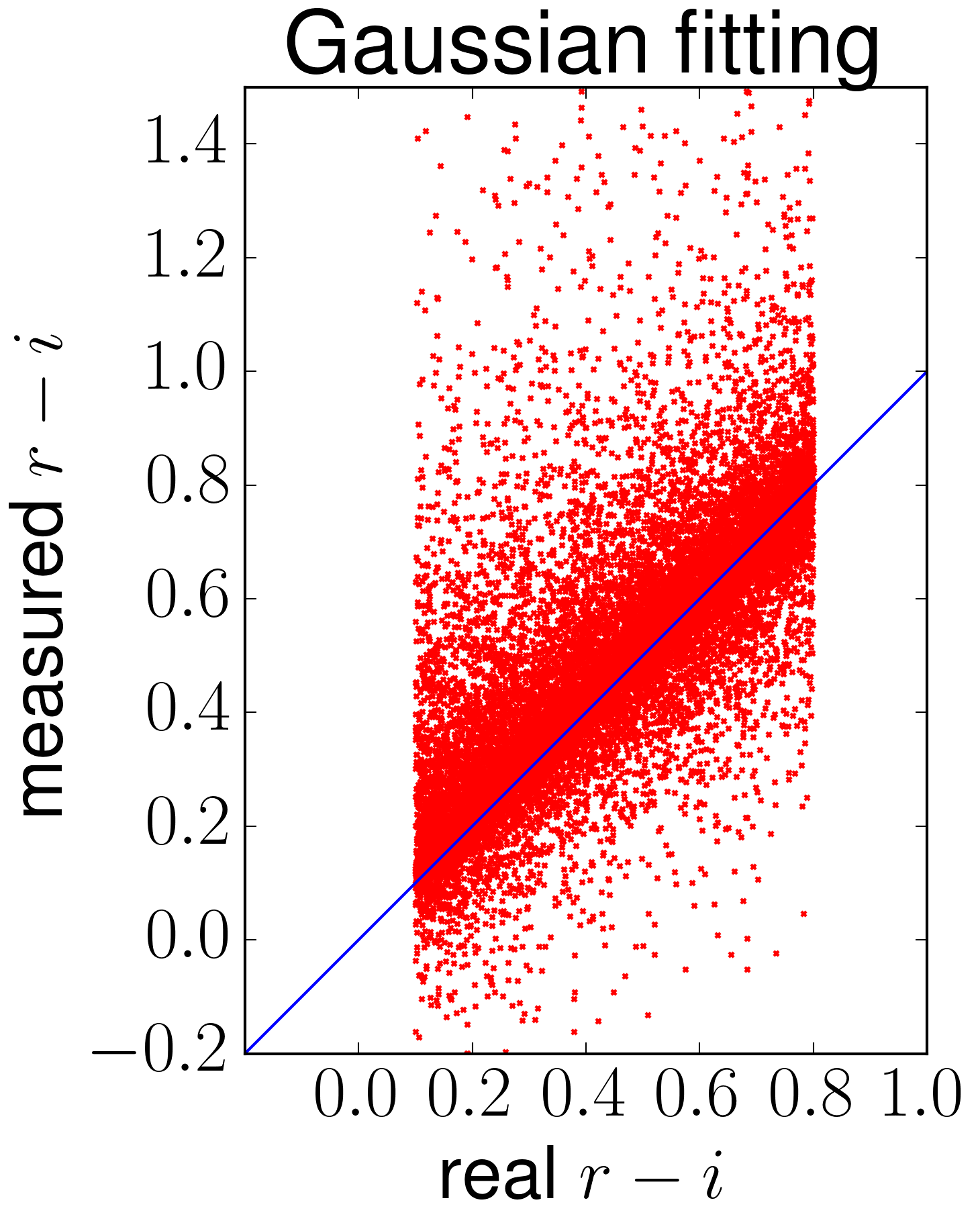}
\end{minipage}
    \caption{Comparison of the real and measured $g-r$ (two left panels) and $r-i$ colours (two right panels) for the mock sample of simulated squares with fixed colours. In each pair, the left panel shows the colour obtained by the linear correlation method and the right panel shows the results obtained by the Gaussian fitting (see the main text). The blue line marks the one-to-one correspondence.}
    \label{fig:com_methods}
\end{figure*}

\begin{figure}

\begin{center}
\includegraphics[width=0.35\textwidth]{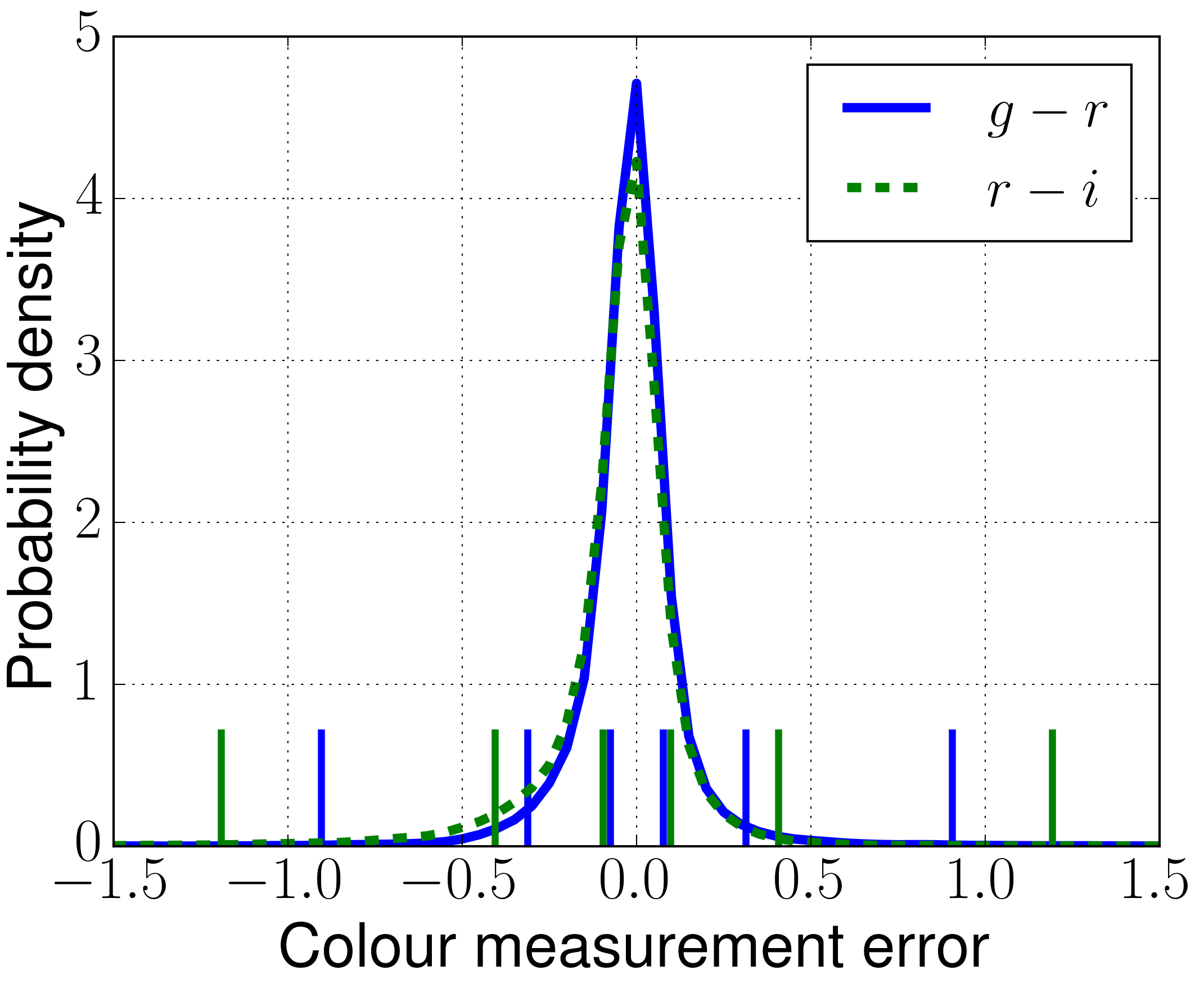}

    \caption{Probability density function for true minus measured colours obtained via Gaussian fitting for simulated filaments. Blue and green vertical lines mark one, two, and three sigma limits for the corresponding distributions.}
    \label{fig:col_diff}
\end{center}    
\end{figure}

\begin{figure}
\begin{center}
    
\begin{minipage}{0.4\textwidth}
\includegraphics[width=0.99\textwidth]{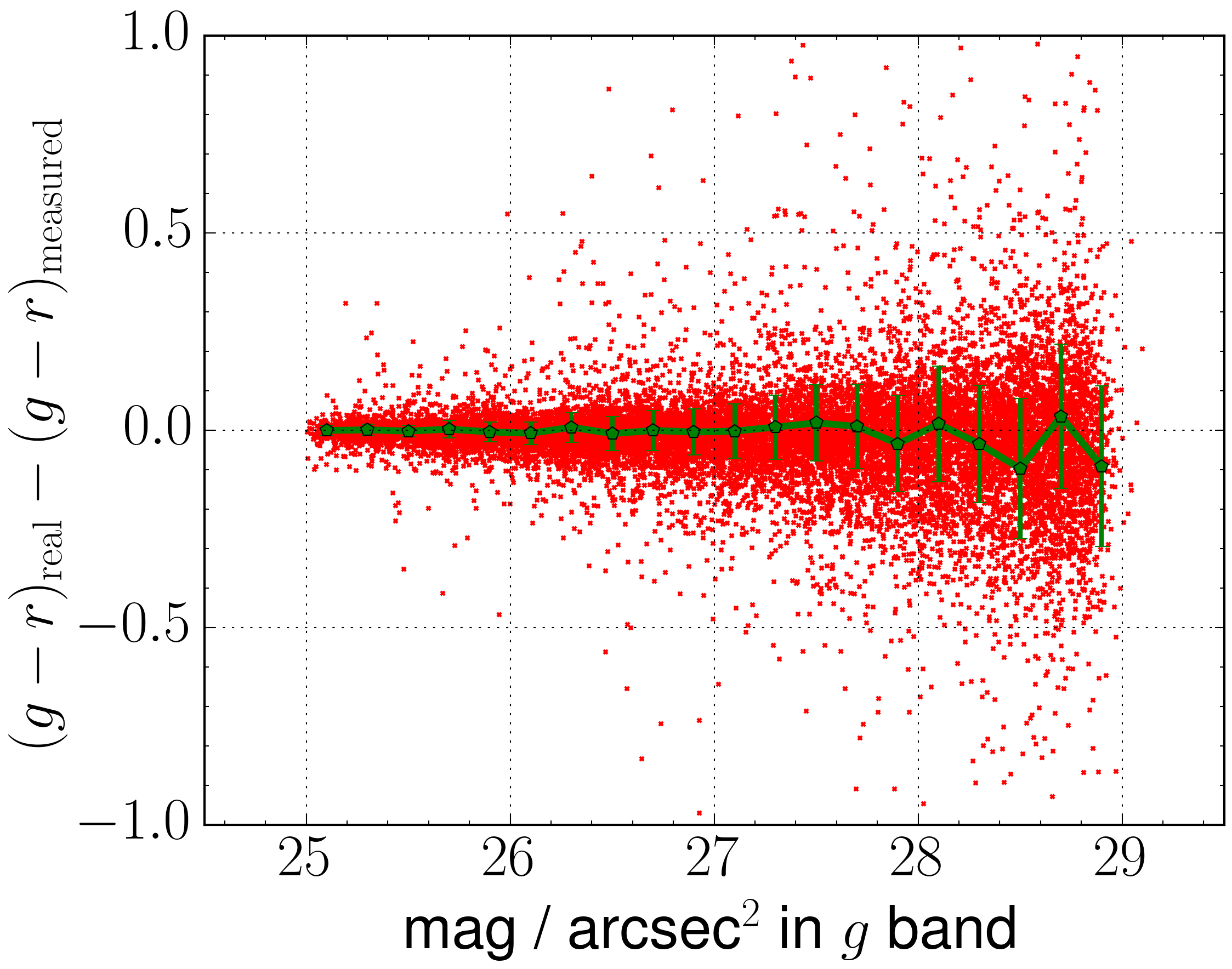}
\end{minipage}\\
\begin{minipage}{0.4\textwidth}
\includegraphics[width=0.99\textwidth]{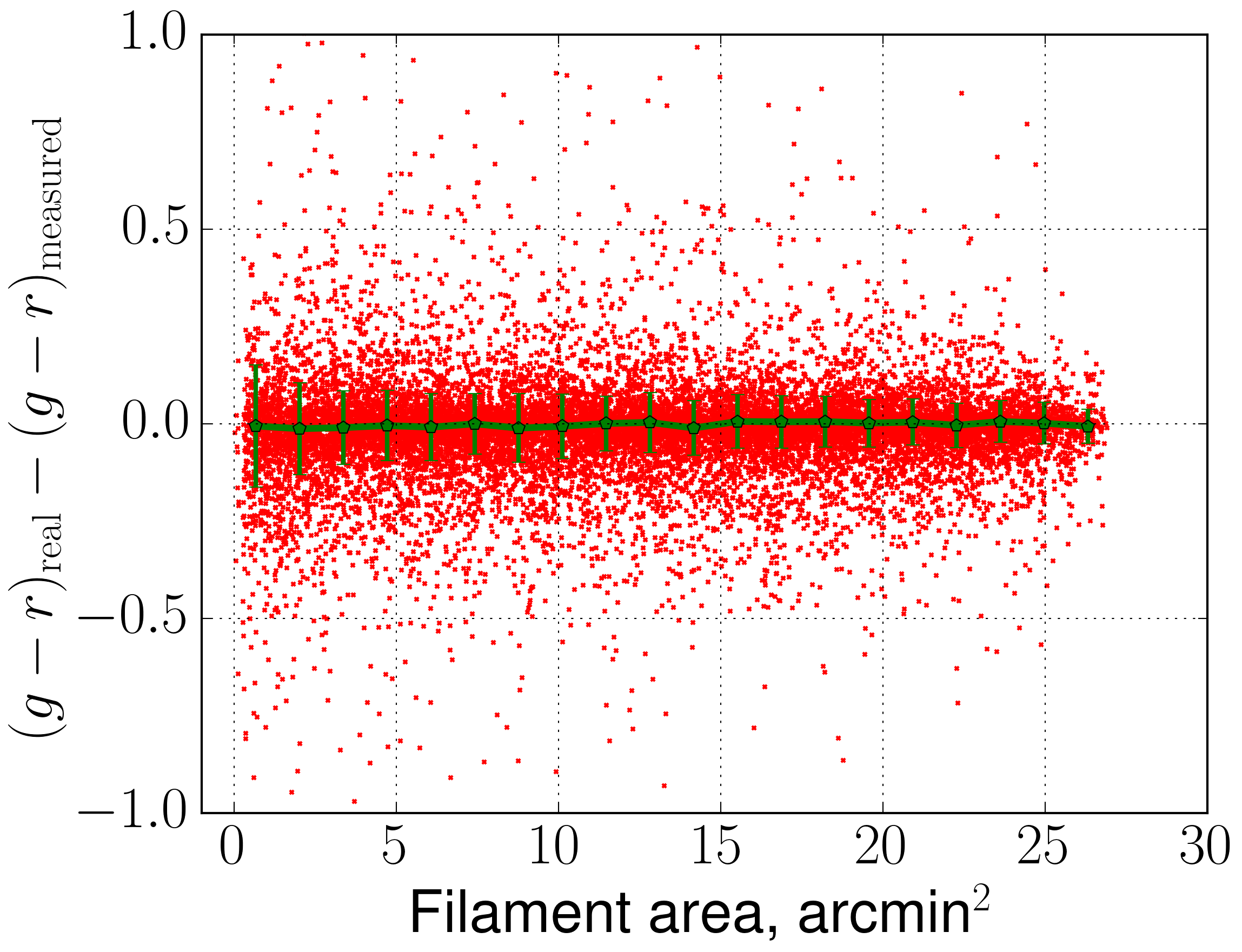}
\end{minipage}%
    \caption{Dependence of the colour measurement error on surface brightness of the \textcolor{black}{simulated} filaments in the $g$ band (top panel) and their area (bottom panel). In both plots, green lines mark the location of the most probable value, while the error bars correspond to 1$\sigma$ limits.}
    \label{fig:col_mu_area}
    
\end{center}
\end{figure}
There is a list of factors that can strongly affect the results of the colour measurements for individual filaments. First of all, at the considered level of surface brightness, the noise can strongly affect the distribution of fluxes. Moreover, the noise also has its own colour properties (due to the differences in band depth), and there is a possibility that the measured colours simply reflect the colours of the noise. Secondly, some other factors are likely to contribute to the measured colours, such as an inaccurate subtraction of the scattered stellar light in the case of very bright stars or the existence of the so-called ``hot'' pixels, which contain emission of some bright, yet poorly resolved sources. Another crucial factor is the sky subtraction, which creates background fluctuations affecting the photometry of extremely low surface brightness sources. How all these factors cumulatively affect the dust colours is hard to estimate analytically. Therefore, to estimate the impact of all these factors, we carried out a series of mock simulations. The general idea of the simulations is to inject an artificial source with a priori known colours into Stripe 82 data and re-measure its colours in a realistic environment where the source is affected by noise, residues of stellar light subtraction, etc. \textcolor{black}{Throughout the present work, we used several types of simulations that differ in the setup of physical parameters. To facilitate the reader, we listed the details of all simulations in Table~\ref{tab:sim}. In this particular section, we discuss the results only for two of them, which are dedicated to study how colour measurement procedures work for individual filaments in general. The respective simulations are labelled as \texttt{S1} and \texttt{S2} in the table. The rest will be discussed below in Section~\ref{sec:res}.}
\par 
The general setup all simulations follow includes the following steps:
\begin{enumerate}
    \item First, we prepare a sample of mock filaments with random sizes, surface brightnesses, and optical colour values. Below we discuss the results for two types of samples, one with a real-like distribution of filament sizes and brightnesses (\textcolor{black}{\texttt{S3} and \texttt{S4}}, Section~\ref{sec:res}), and the other with a uniform distribution of these properties (\textcolor{black}{\texttt{S1} and \texttt{S2}}, this Section). The distribution of the colours for real filaments is actually unknown and, therefore, we choose the colours uniformly in some predefined range. \textcolor{black}{We consider a rather wide range of colours, for example, from 0.1 to 0.8 for $g-r$ (see Table~\ref{tab:sim}), because, as we show below, real filaments' colours also tend to have a wide spread.} To simplify the analysis, each filament has a square-like shape. As for the areas, we originally selected them  in the following range: from $144$ arcsec$^2$ (four pixels) up to $10^5$ arcsec$^2$ ($\approx$ 27 arcmin$^2$). But due to the mask, some pixels are cut, and, therefore, the actual area of each filament slightly varies from the predetermined set of values. 
    \item Secondly, we inject a square  with the selected size, surface brightness, and colours from the prepared sample into Stripe 82 data. The centre of the square is chosen randomly, that is, the square is located at random point of Stripe 82. Next, we add some flux values in each band to all the pixels within the square area. The values are selected so as to have some average value corresponding to the initially selected surface brightness value, with a small variance. The variance is the same for all filaments and is equal to 20 counts (in the $g$ band). The value is close to the typical spread of values for real filaments. For $r$ and $i$ bands, the fluxes are determined from the fluxes in the $g$ band, assuming the constant value of $g-r$ and $r-i$ optical colours over the square. For each band, we also modify the distribution of fluxes to take into account the Poisson noise from the source. To calculate the number of events for the Poisson statistics, we assume the following $gain$ values: 3.85, 4.735, and 5.15 for $g$, $r$, and $i$ bands, respectively. These values are obtained by averaging $gain$ values for different $camcol$ parameters of SDSS imaging camera.
    \item Thirdly, we measure the colours in exactly the same way as we do for real filaments (real cirrus filaments are also masked for the purpose of simulations). For measurements themselves, we adopt two different approaches (see Fig.~\ref{fig:fil_measurement}): a classical linear correlation method~\citep{Guhathakurta1989, 2010ApJ...723.1549S, 2014ApJS..213...32M, Roman_etal2020} and the method suggested by~\cite{Roman_etal2020}, which is based on the analysis of colour distribution of individual pixels. We discuss the applicability of both methods to the measurement of individual filaments below.
    \item Finally, we compare the measured colours with their true values, and check what factors are important for reliable colour measurements.
    
\end{enumerate}
\par 
Here we briefly discuss the details of the two adopted methods of colour measurement. 
\par 
The essence of the first method is the linear correlation between the fluxes in different bands. While fitting the linear dependence to the distribution of fluxes, for example, in $g$ and $r$ bands, one finds a linear coefficient, which can be translated into the corresponding $g-r$ colour value~(see Fig.~\ref{fig:fil_measurement}, two left panels). While this method is commonly used for colour measurement, the resulting colours obtained using this method can be significantly affected by the noise in a low surface brightness regime as we show below.
\par 
The second method, introduced in~\cite{Roman_etal2020}, assumes that, for a particular cloud, the real distribution of dust colours  should be close to Gaussian, and the position of the Gaussian maximum should correspond to the actual colour of the cloud. The noise contributions are accounted for in this approach through the simultaneous fitting of the Lorentzian function, which describes the noise, and Gaussian function, which describes the distribution of real dust colours. Testing how this approach works for different filaments, for which the number of pixels is considerably smaller than in~\cite{Roman_etal2020}, we found that a simultaneous fitting of Gaussian and Lorentz functions with a full set of free parameters can lead to degenerate results, or there can be a set of close solutions that have different colour values. The problem can be solved by a manual analysis of the fitting results and rejecting non-physical results, but an identification of the fitting failure for a large number of the filaments is a complex problem. Thus, we use a more constrained approach, omitting the Lorentz part and fitting only the Gaussian part. 
We justify such a simplification based on our results from simulations presented below. We should also note that, originally, we tried to estimate the Lorentz function parameters from the layer of the pixels that are close to the filament, but which do not include it. Then we tried to fit the Gaussian function along with the Lorentz function, fixing some parameters for both functions (like the Lorentz peak location and its scale, and the Gaussian amplitude). We found that, for such a setup, the resulting colours are very close to the case when we fit only a Gaussian part. 
\par
Fig.~\ref{fig:com_methods} presents the distribution of real versus measured $g-r$ and $r-i$ colours for both approaches discussed above. The values were obtained by measuring the colours of $2 \cdot10^3$ squares with uniformly distributed colours from $0.1$ to $0.8$ for both $g-r$ and $r-i$ and surface brightnesses ranging from 25 mag arcsec$^{-2}$ to 29 mag arcsec$^{-2}$ in the $g$ band (\textcolor{black}{see \texttt{S1} simulation from Table~\ref{tab:sim}}). 
\par 
As can be clearly seen from Fig.~\ref{fig:com_methods}, for both $g-r$ and $r-i$, there is no consistency between real and measured colours if the colours are obtained using the fluxes correlation method ($r-i$ colours are systematically greater on average). At the same time, there is much desired one-to-one correspondence for most of the filaments if we measure the colours by fitting the Gaussian function to the colour distribution. Our results show that the mode of the colour distribution is a more stable parameter than the coefficient of the linear correlation in the case of a significant noise contribution to the fluxes. We also note that we apply linear correlation method without introducing some limiting surface brightness value like in~\cite{Roman_etal2020}, since the vast majority of our filaments have a very low surface brightness and insufficient to make such cuts. Based on the results, we conclude that the linear correlation method is unreliable for colour measurement of individual filaments, while our second method allows one to retrieve the actual colours for most of the clouds.
\par 
The probability density function of the true minus measured colours obtained via Gaussian fitting is presented in Fig.~\ref{fig:col_diff}. We also mark three limits: 0.08,  0.31, 0.90 for $g-r$ and 0.10, 0.40, 1.20 for $r-i$. Within these limits lie $68.27\%$, $95.45\%$, and $99.73\%$ of all filaments, respectively. The values give a qualitative understanding of what errors one should expect from the measurement of real filaments. It also shows that, unfortunately, for individual filaments the errors can be quite large, up to $1.0$. With such an error, any physical comparison with other sources is essentially meaningless. At the same time, if one considers a large sample, there should be many filaments for which the colours are measured with an acceptable error of $0.1-0.2$ (in absolute units, that is, mag). We exploit this fact below when interpreting the observed distribution of the colours of real filaments.

\par
To facilitate future studies of dust colour over small spatial scales, we verify how the difference between true and measured colours depends on filament area and surface brightness. Fig.~\ref{fig:col_mu_area} presents the mentioned dependencies for $g-r$ (for $r-i$ all presented dependencies are qualitatively the same). As one naturally expects, the surface brightness is important and, as the surface brightness increases, the colour measurement error decreases. \textcolor{black}{As can be seen from the figure, for most filaments of 26 mag/arcsec$^2$ and brighter, the error of colour measurement is smaller than 0.05. Such bright filaments are most likely to be identified by their true colours in Stripe 82. For dim filaments, the error increases rapidly after 26.5 mag/arcsec$^2$, reaching about 0.10 at 27 mag/arcsec$^2$ and about 0.20 at 28 mag/arcsec$^2$. As mentioned above, such a large error makes it hard to distinguish the filaments from other sources by their colours in practice. As shown in the lower panel of Fig.~\ref{fig:col_mu_area},} increasing the area of the filaments certainly helps too, although the effect is not that prominent when compared to the case of surface brightness. 
\par
\begin{figure}
\begin{center}
    
\begin{minipage}{0.4\textwidth}
\includegraphics[width=0.99\textwidth]{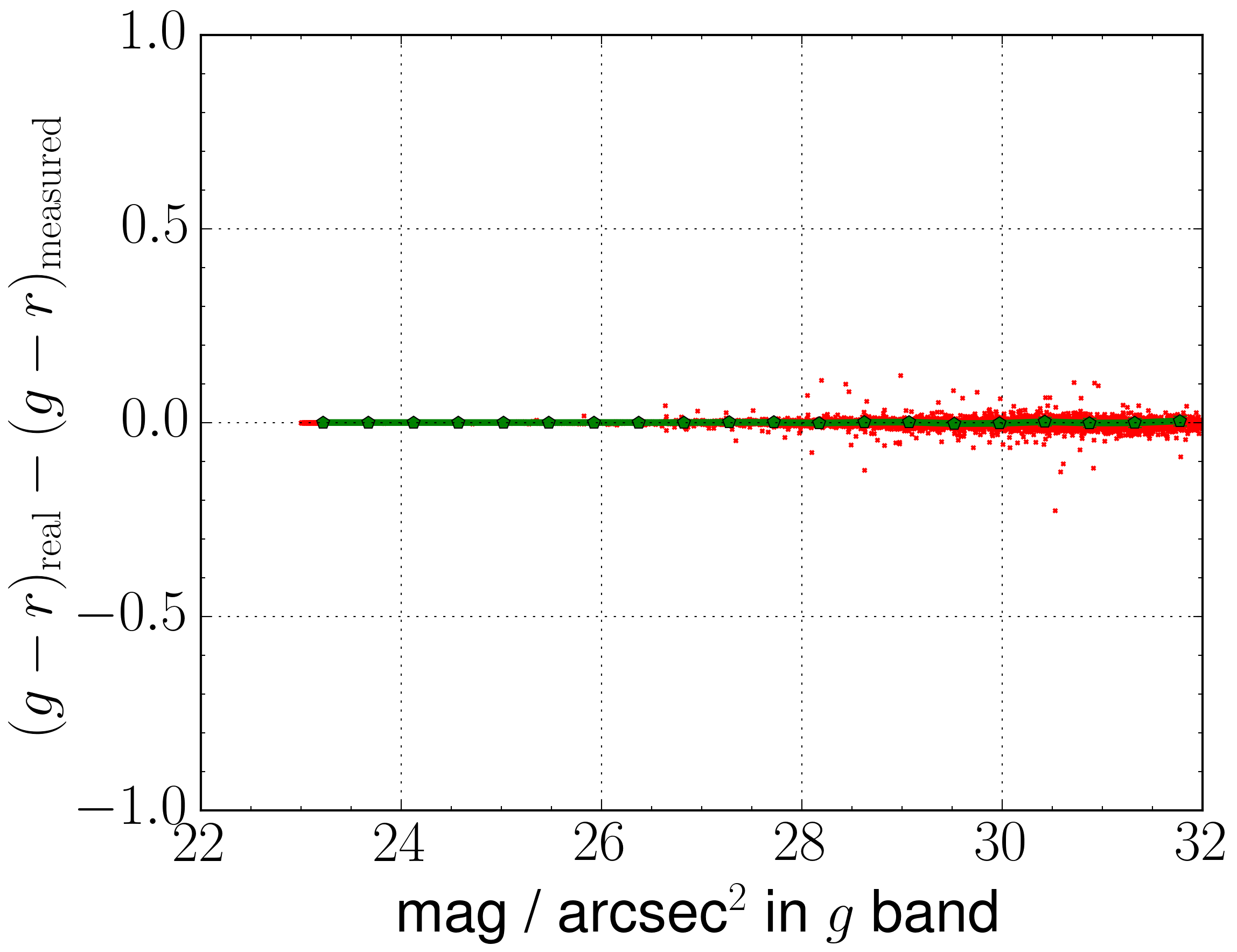}
\end{minipage}\\
\begin{minipage}{0.4\textwidth}
\includegraphics[width=0.99\textwidth]{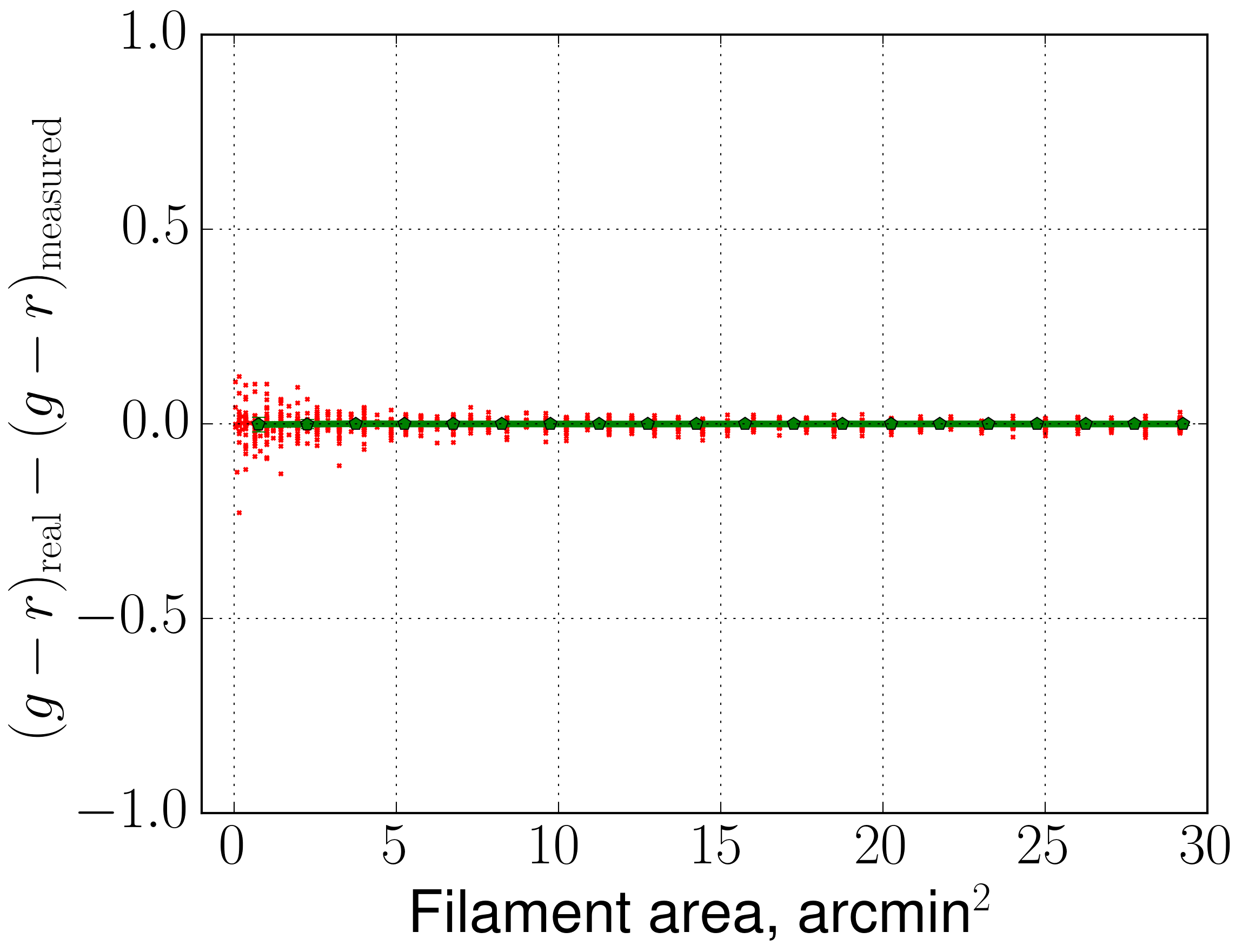}
\end{minipage}%
    \caption{Same as Fig.~\ref{fig:col_mu_area}, but for \textcolor{black}{simulated} filaments with wider ranges of area and surface brightness and inserted into a clean field, where only Gaussian noise presents without other sources, mask, etc. }
    \label{fig:col_mu_area_clear}
    
\end{center}
\end{figure}

As an additional test, we performed similar simulations inserting mock filaments into the artificial field with only noise present (no other sources, no mask, etc). \textcolor{black}{This simulation is labelled as \texttt{S2} in Table~\ref{tab:sim}. The only difference between this simulation and the previously considered \texttt{S1} is the background into which the squares are injected. In the case of \texttt{S1}, the background is Stripe 82 fields, while for \texttt{S2}, the background consists only of artificially created noise.} The noise characteristics were selected to reflect $g$ and $r$ Stripe 82 depth limits, which are $\mu_{g, lim}=29.2$ mag arcsec$^{-2}$ and $\mu_{r,lim}=28.7$ mag arcsec$^{-2}$, respectively, measured over boxes of 10 arcsec. As can be seen from Fig.~\ref{fig:col_mu_area_clear}, in the ``ideal'' situation with only the noise present, one should retrieve the colours of the filaments with a much higher degree of accuracy than the filaments from the actual Stripe 82 data show. 
\par 
\textcolor{black}{The top panel of Fig.~\ref{fig:col_mu_area_clear} shows that the error in colours of very faint filaments with $\mu\approx 31$ mag arcsec$^{-2}$ is small ($\lesssim 0.1$ mag), despite the fact that such surface brightnesses are clearly below the surface brightness limits introduced earlier. There is actually no contradiction because the limiting surface brightness values are those typically defined in 10x10 square arcseconds. However, the simulated filaments have areas that are orders of magnitude larger than the area in which the surface brightnesses limits are defined. This is clearly seen in the bottom panel of Fig.~\ref{fig:col_mu_area_clear}, where the main limiting factor is in fact the area of the filaments. Since the filaments have such a large size, the limiting surface brightnesses in this extremely large area range are very high. For example, the limiting surface brightness of SDSS Stripe 82 at 10x10 arcsec$^2$ is 29.2 mag arcsec$^{-2}$ in the $g$ band, which translates to 31.2 mag arcsec$^{-2}$ at 1x1 arcmin$^2$, typical explored area of the filaments. This surface brightness value is at the upper limit of the magnitudes considered in our tests.}
\par 
\textcolor{black}{Overall, the comparison of Fig.~\ref{fig:col_mu_area} and Fig.~\ref{fig:col_mu_area_clear} indicates that a significant portion of error in colour measurement comes from the background into which the squares are injected. Ideally, if the background is processed accurately, this should not be the case. This indicates that the data processing itself is a very important factor for colour measurements.} 
There are many steps to it, including those carried out not in the present work, like flatfielding and sky subtraction. It would be an interesting problem to consider how much each of these steps contribute to the overall error, but we do not go further in this direction in the present work.

\section{Results}
\label{sec:res}
\begin{figure}
	
    \begin{center}
        \includegraphics[width = 0.38\textwidth]{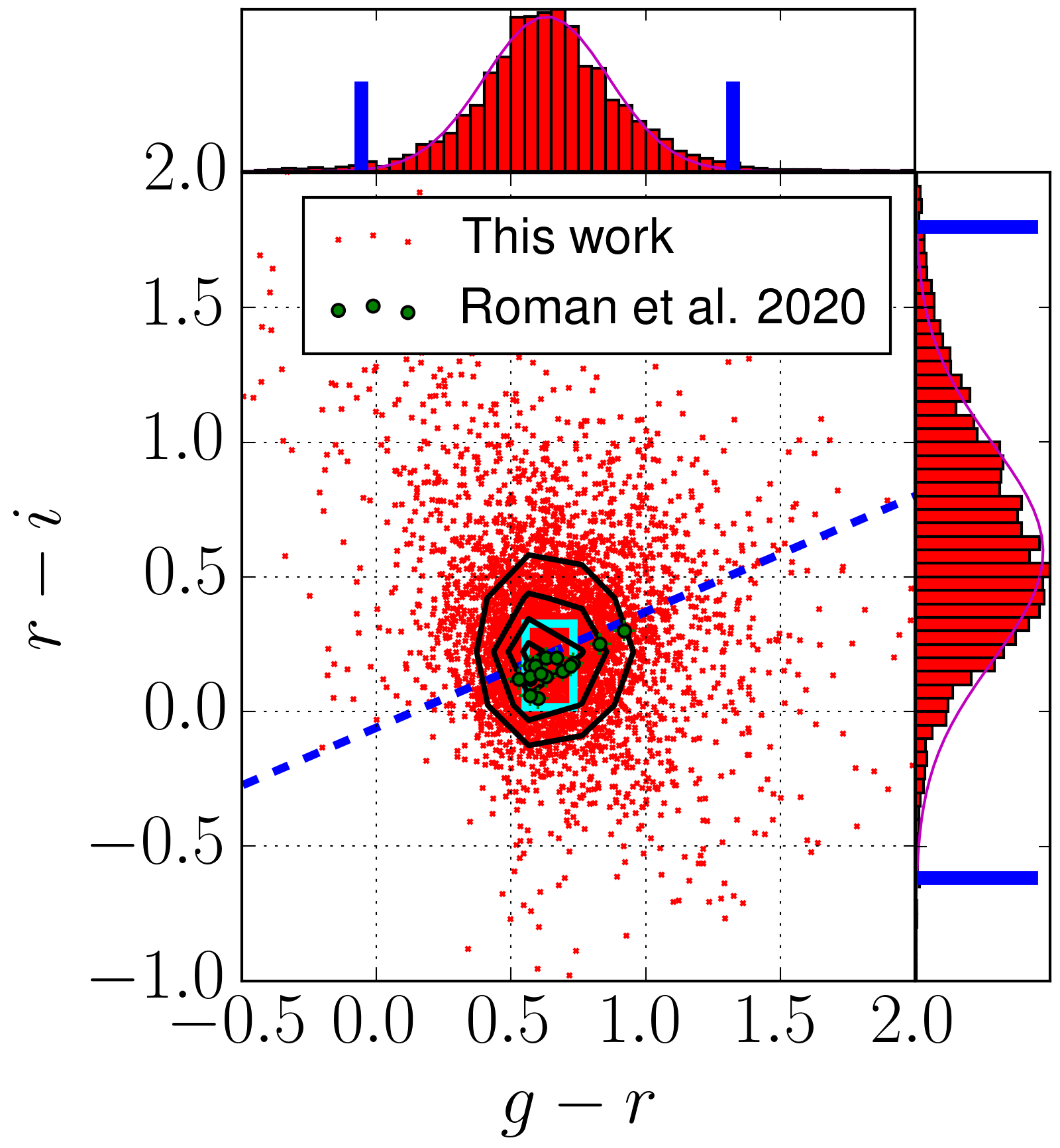}
    \end{center}
    \caption{Distribution of the cirrus colours (red points) from the present work and~\protect\cite{Roman_etal2020} (green points). The blue line $(r - i) = 0.43 \times (g - r) - 0.06$ should separate the cirrus colours and the colours of extragalactic sources, as suggested by~\protect\cite{Roman_etal2020}. The light blue rectangle marks the estimated dispersion of true cirrus colours (see the main text). Blue bars mark 3$\sigma$ limits of the corresponding distributions.}
    \label{fig:colours_6arcsec}
\end{figure}
The 2D distribution of the $g-r$ and $r-i$ colours for the filaments identified in Stripe~82 are presented in Fig.~\ref{fig:colours_6arcsec}. The colours are measured using the Gaussian fitting method described in the previous section.  In the same figure, we also depict the results of~\cite{Roman_etal2020} for their sixteen fields, and the line $(r - i) = 0.43 \times (g - r) - 0.06$, which should separate the colours of cirrus filaments and other extragalactic sources, as~\cite{Roman_etal2020} suggested. In subpanels of the figure, we plot individual distributions of the colours, their respective Gaussian approximations (magenta lines), and 3$\sigma$ limits (thick blue rectangles). From the figure, one can indicate two important properties of the colour distribution. First, there is a peak of the density contours at about $g-r\approx0.6$ and $r-i\approx0.2$. Secondly, there is a large spread of values in both $g-r$ and $r-i$ colours, $\sigma$ is about 0.3 for $g-r$ and 0.4 for $r-i$. 
\par 
We note that peak locations of the 1D distributions, displayed in the side panels of Fig.~\ref{fig:colours_6arcsec}, can be somewhat misleading. For example, the Gaussian of $r-i$ colours has the peak located at $r-i=0.59$. This is significantly greater than the corresponding $r-i\approx0.2$ of the 2D peak. The reason for this is that, for different $r-i$ values, $g-r$ values are also distributed differently. In the upper part of the plot (\mbox{$r-i\gtrsim$ 0.3-0.4}), $g-r$ colour are distributed sparsely for a fixed value of $r-i$ and, thus, no density peak is observed. For $r-i\lesssim$ 0.3-0.4, $g-r$ are clustered very closely and there is a density peak. For a fixed value of $r-i$ (for example, for $r-i\approx0.6$ and $r-i\approx0.4$), the total number of filaments  is nearly the same in both cases.
\par 
\begin{figure}
    \begin{center}
        \includegraphics[width = 0.38\textwidth]{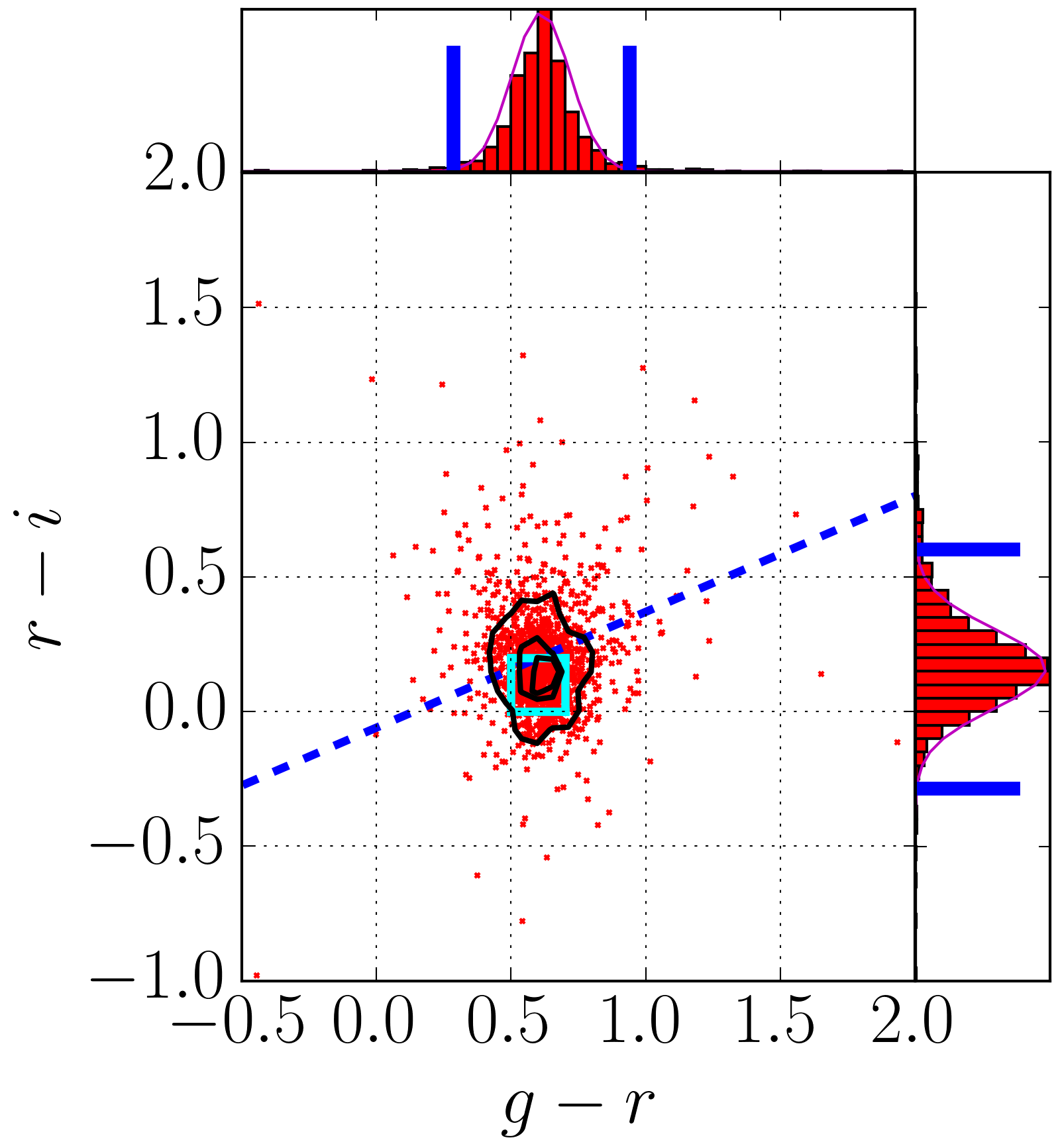}
    \end{center}
    \caption{Same as Fig.~\ref{fig:colours_6arcsec}, but for a sample of simulated filaments, the true colours of which are uniformly distributed within the area marked by a light blue square.}
    \label{fig:fil_sim_ex}
\end{figure}

\begin{figure*}
    \centering
    \begin{minipage}{0.25\textwidth}
    \includegraphics[width=0.99\textwidth]{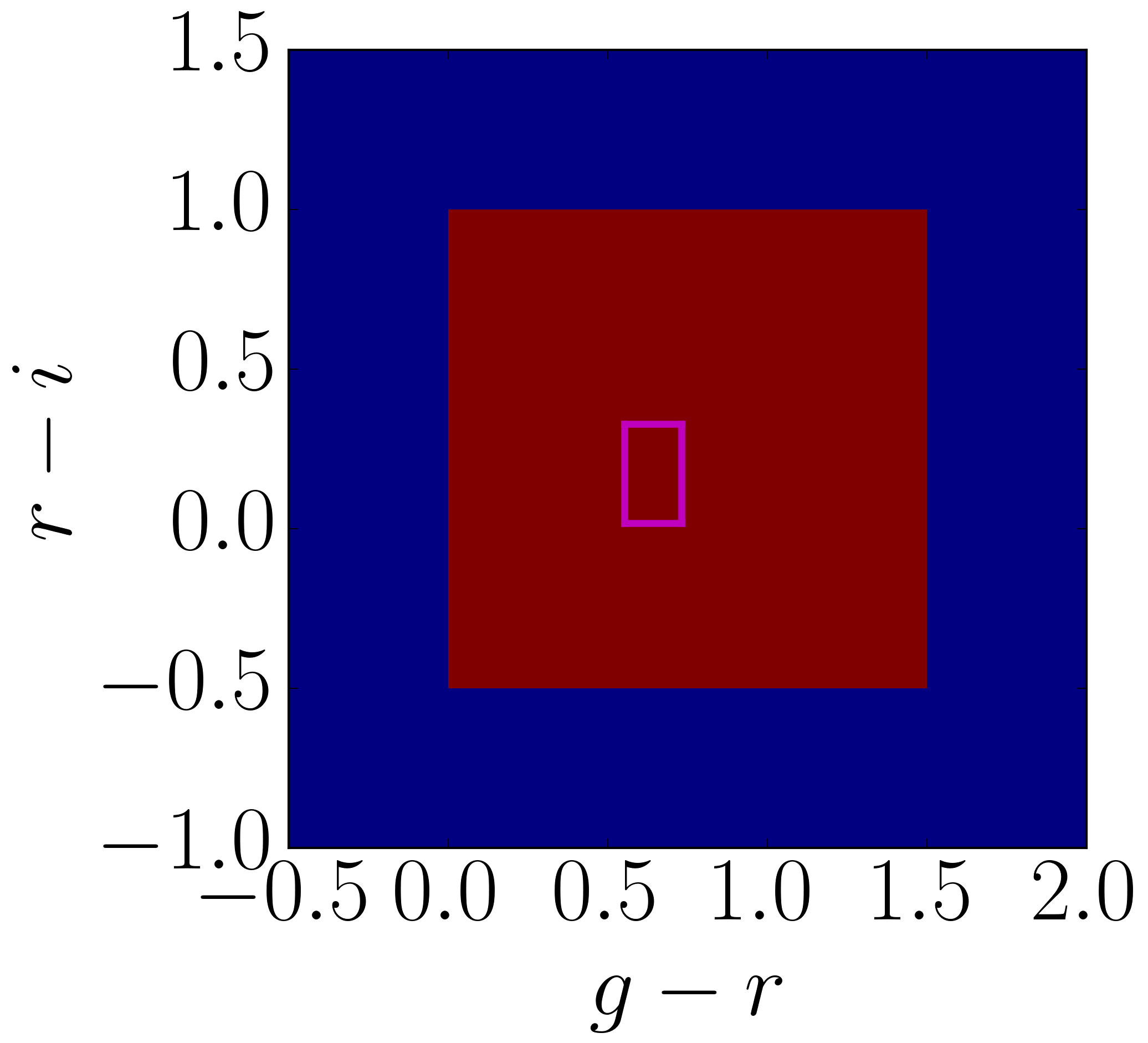}
    \end{minipage}%
    \begin{minipage}{0.25\textwidth}
    \includegraphics[width=0.99\textwidth]{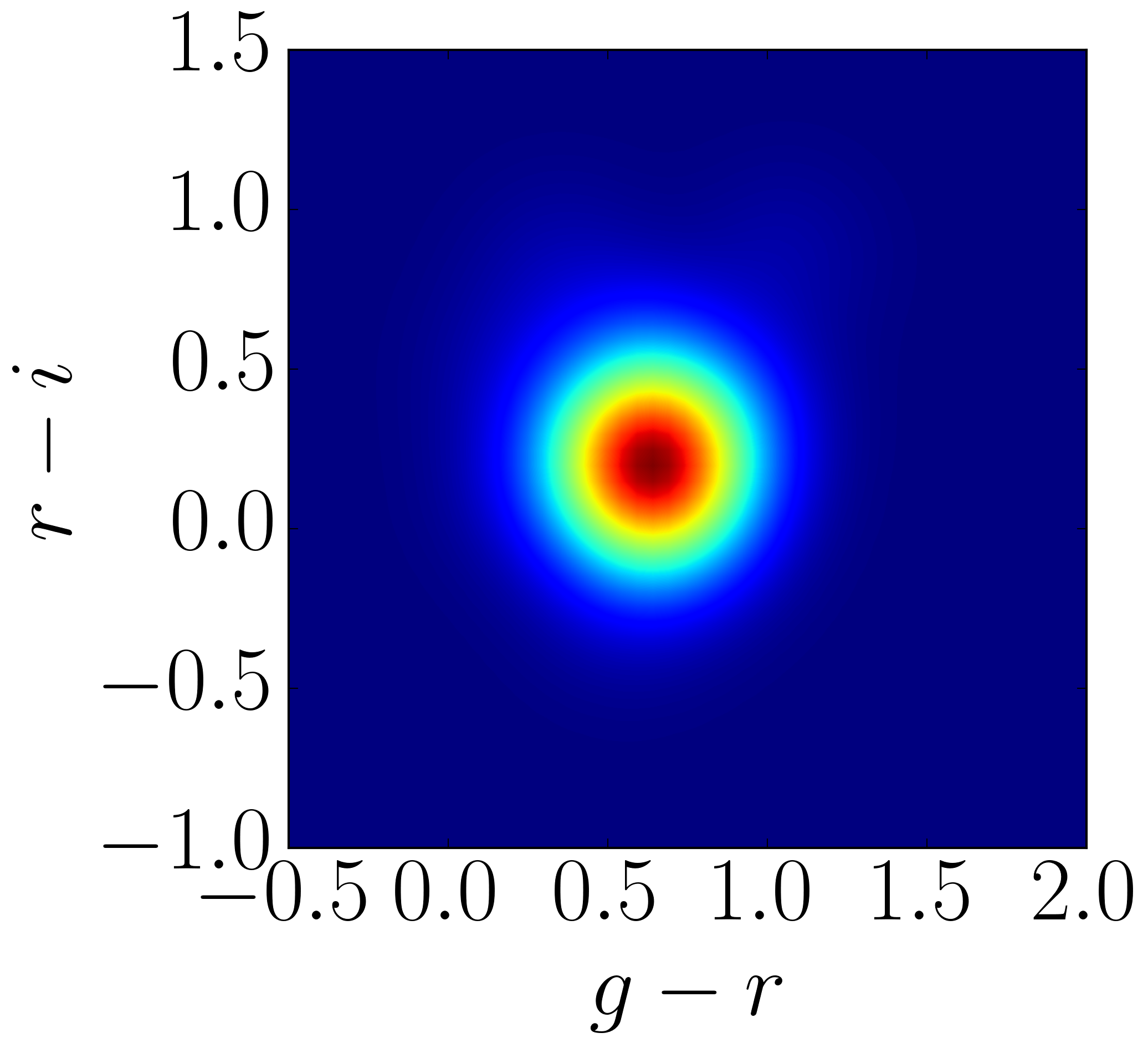}
    \end{minipage}%
    \begin{minipage}{0.25\textwidth}
    \includegraphics[width=0.99\textwidth]{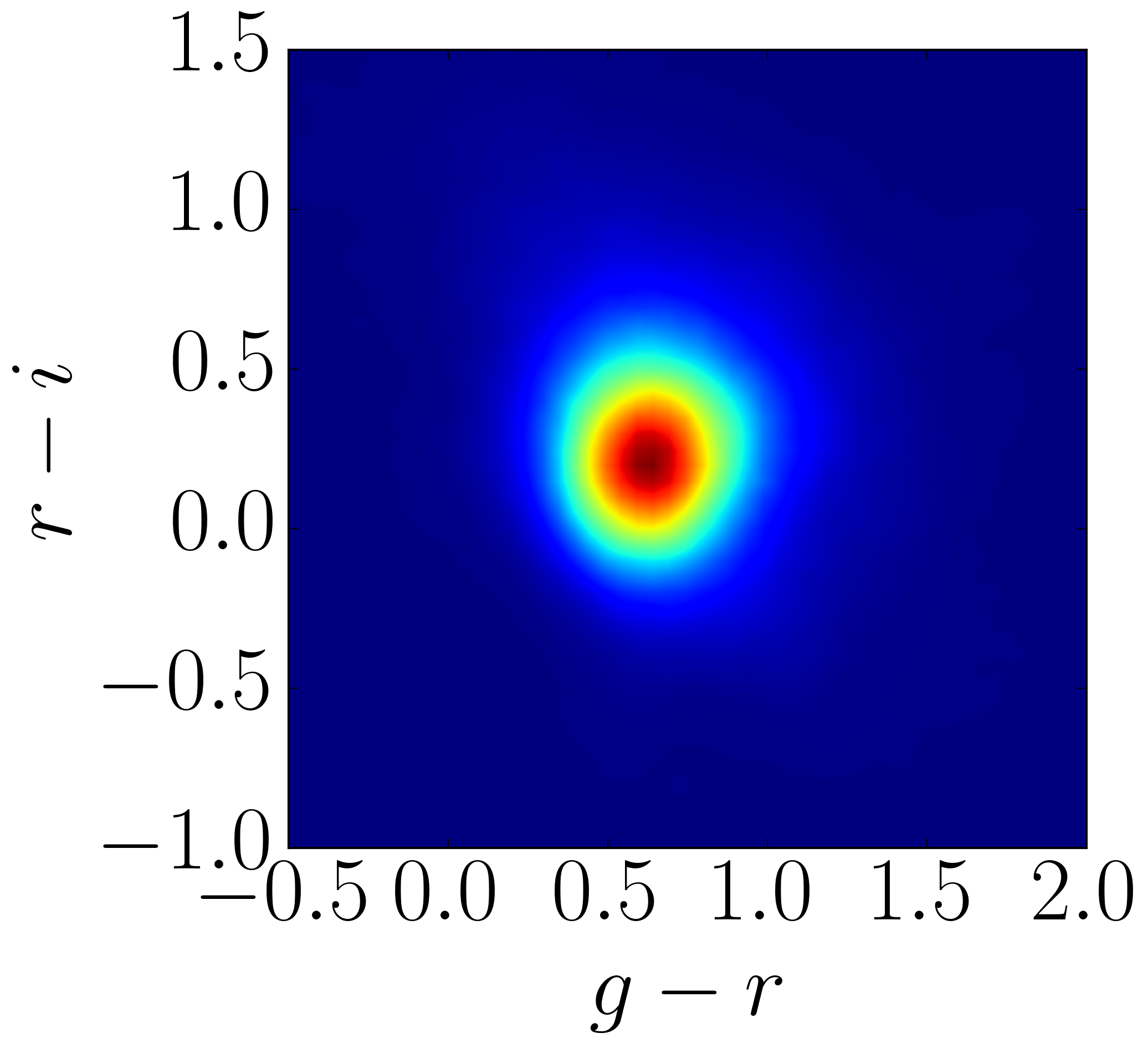}
    \end{minipage}%
    \begin{minipage}{0.29\textwidth}
    \includegraphics[width=0.99\textwidth]{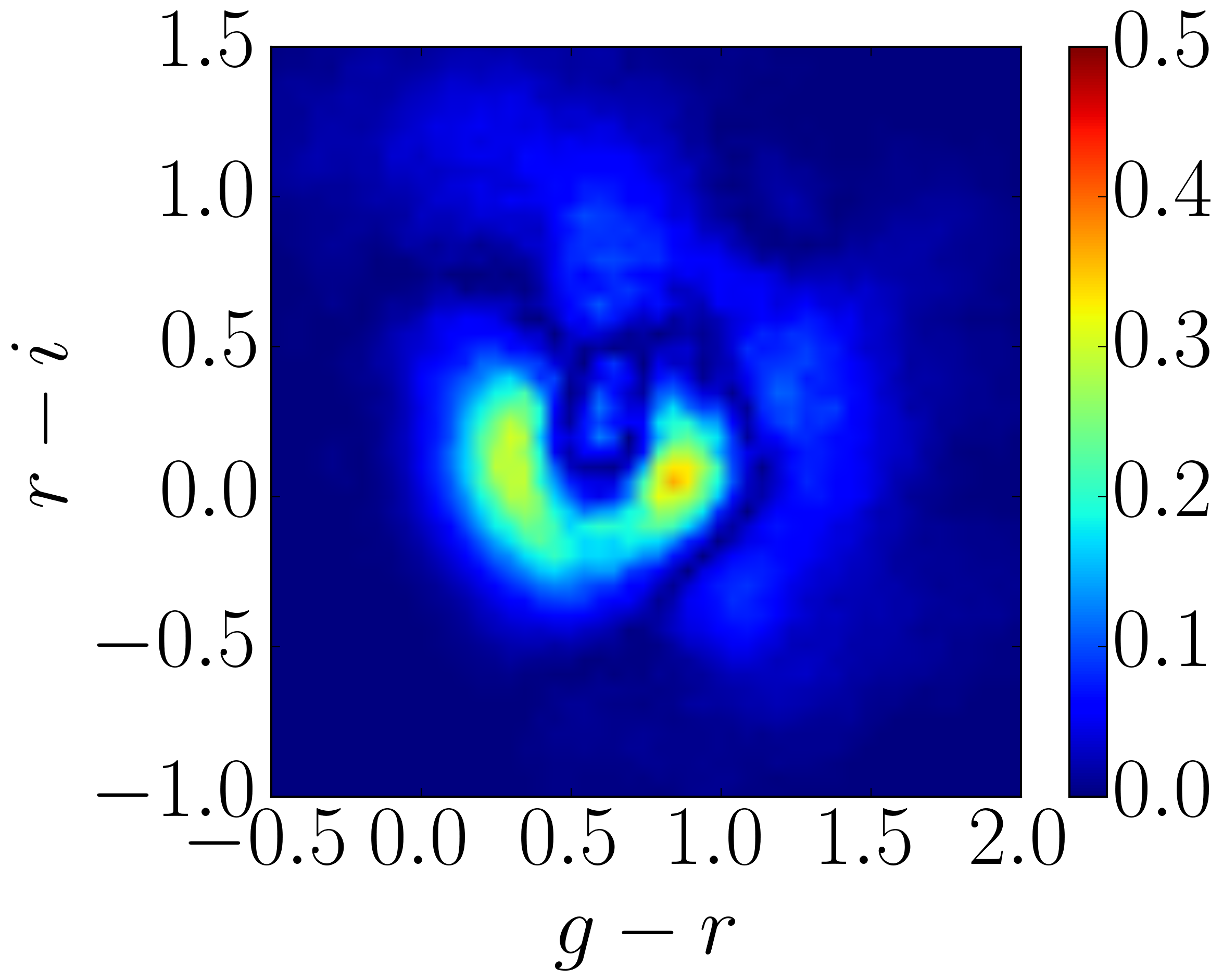}
    \end{minipage}
    \caption{Finding the dispersion of the colours for the real filaments. Leftmost panel: an initial uniform distribution of colours for mock filaments and a selection of a smaller area~(the magenta rectangle) to find an optimal colour range for the real filaments. Second left panel: a smoothed map of the colour distribution produced by the filaments from the magenta square marked in the leftmost panel. Third left panel: a similar smoothed map for the real filaments. Rightmost panel: the residue between  the map for real filaments and the best-fit map for the mock filaments. The residual values correspond to the difference between the probability density functions, obtained by properly \textcolor{black}{normalizing} both maps. Thus, the units of the colourbar are the units of the probability density function, $1/(mag \times mag)$.}
    \label{fig:fil_optimal}
\end{figure*}

In the previous section, we concluded that the colours of dim filaments are significantly affected by various contaminating factors (noise, masking residues, etc). Therefore, it is only natural to ask to what degree the observed spread of the values corresponds to the real dispersion of the cirrus colours. To answer this question, we consider a sample of mock filaments (squares) with the sizes and surfaces brightnesses distributed according to the distribution of these properties for real filaments, presented in the left panel of Fig.~\ref{fig:correlations} (in contrast to a sample considered in Section~\ref{sec:colour_measurement}, where these properties were uniformly distributed). \textcolor{black}{This simulation is labelled as \texttt{S3} in Table~\ref{tab:sim}.} We carry out the simulation for such a sample in the same manner as it was done in~Section~\ref{sec:colour_measurement} comparing real and measured colours. 
For the original colours, we select a uniform distribution within the following limits: $0.5\leq g-r\leq0.7$ and $0.0\leq r-i\leq0.2$. Fig.~\ref{fig:fil_sim_ex} shows the resulting colours for the sample. The light blue square marks the limits of the original colours. The obtained distribution is qualitatively similar to that for the real filaments. Again, there is a clear density peak at  $g-r=0.60$, $r-i=0.10$ (average of the originally selected values) and rather extended wings (see blue rectangles). These wings lie outside of the square of the original colours. This means that the wings arise due to contamination factors and, therefore, do not reflect the real dispersion of the originally selected colours. For the real filaments, we assume that the situation should be qualitatively the same. The large spread of colours displayed in Fig.~\ref{fig:colours_6arcsec} should be due to contamination factors discussed in Section~\ref{sec:colour_measurement}, and does not reflect the real difference in the cirrus colours. The real variation of cirrus colour should manifest itself in the structure of the densest part of the distribution. 

Since the distributions for real and simulated filaments are still qualitatively similar (the dense part plus the wings), one can try to identify the real dispersion of cirrus colours applying some kind of ``deconvolution'' procedure. We use the following approach. First, we expand our simulations and consider a sample of mock filaments with the colours initially distributed uniformly in a wide range, $0.0\leqslant g-r\leqslant1.5$ and $-0.5\leqslant r-i\leqslant1.0$ (\textcolor{black}{simulation \texttt{S4} in Table~\ref{tab:sim}}). Then we construct a specific function, the purpose of which is to produce 2D density maps of the filament colours on the $(g-r, r-i)$ plane based on the true colours of the filaments. The details are as follows: 
\begin{enumerate}
    \item First, the function accepts some colour ranges as arguments and finds the filaments in the simulated sample with the original colours within the originally selected limits (Fig.~\ref{fig:fil_optimal}, \textit{leftmost} panel). For simplicity, the selected area has a rectangular shape.
    \item Secondly, the function assesses the measured colours, which differ from thier true colours, and which are distributed in a manner similar to that shown in Fig.~\ref{fig:colours_6arcsec} and Fig.~\ref{fig:fil_sim_ex}. From the distribution, a smooth density profile is created via the kernel density estimation procedure from the \texttt{python} package \texttt{sklearn} (Fig.~\ref{fig:fil_optimal}, \textit{second left} panel). The resolution of the prepared map is 0.05 along both axes. 
    \item Thirdly, we prepare a similar smooth density map for real filaments ~(Fig.~\ref{fig:fil_optimal}, \textit{third left} panel).
    \item At the last step, we find an optimal range of the colours which minimises the sum of square differences between the density map for simulated filaments and the similar map for real filaments. A typical residual map is presented in the rightmost panel of Fig.~\ref{fig:fil_optimal}. 
    
 \end{enumerate}  
 
As a result of the analysis, we find that the closest to real observable distribution is produced by filaments with colours in the following ranges: $0.55 \leqslant g-r \leqslant0.73$ and $0.01\leqslant r-i \leqslant0.33$. We marked this area by a light blue rectangle in Fig.~\ref{fig:colours_6arcsec}. As can be seen, almost all clouds from~\cite{Roman_etal2020} have the colours within these limits, except for two of them, which are outside of the region. Thus, for most filaments from our sample, the colours are consistent with those measured by~\cite{Roman_etal2020}, that is, when the colour is averaged over large spatial areas in Stripe~82. As for criterion~(\ref{eq:Roman}), suggested by~\cite{Roman_etal2020}, the most part of the rectangle is located below the separating line, although there is also a slight area above it. Does this mean that the condition is violated? The correct answer is that, given the accuracy for colour estimation for individual filaments (which should be about 0.1 for most filaments, see Fig.~\ref{fig:col_diff}), it is impossible to say whether this is really the case. Moreover, our approach to identify the real colours assumes that the colours of filaments are distributed uniformly, which is, of course, a massive simplification. Thus, we conclude that  more precise data is required to verify whether condition~(\ref{eq:Roman}), suggested by~\cite{Roman_etal2020}, holds on a spatial scale of individual filaments.

\par
\begin{figure*}
    \centering
    \begin{minipage}{0.45\textwidth}
    
    \includegraphics[width=1.0\textwidth]{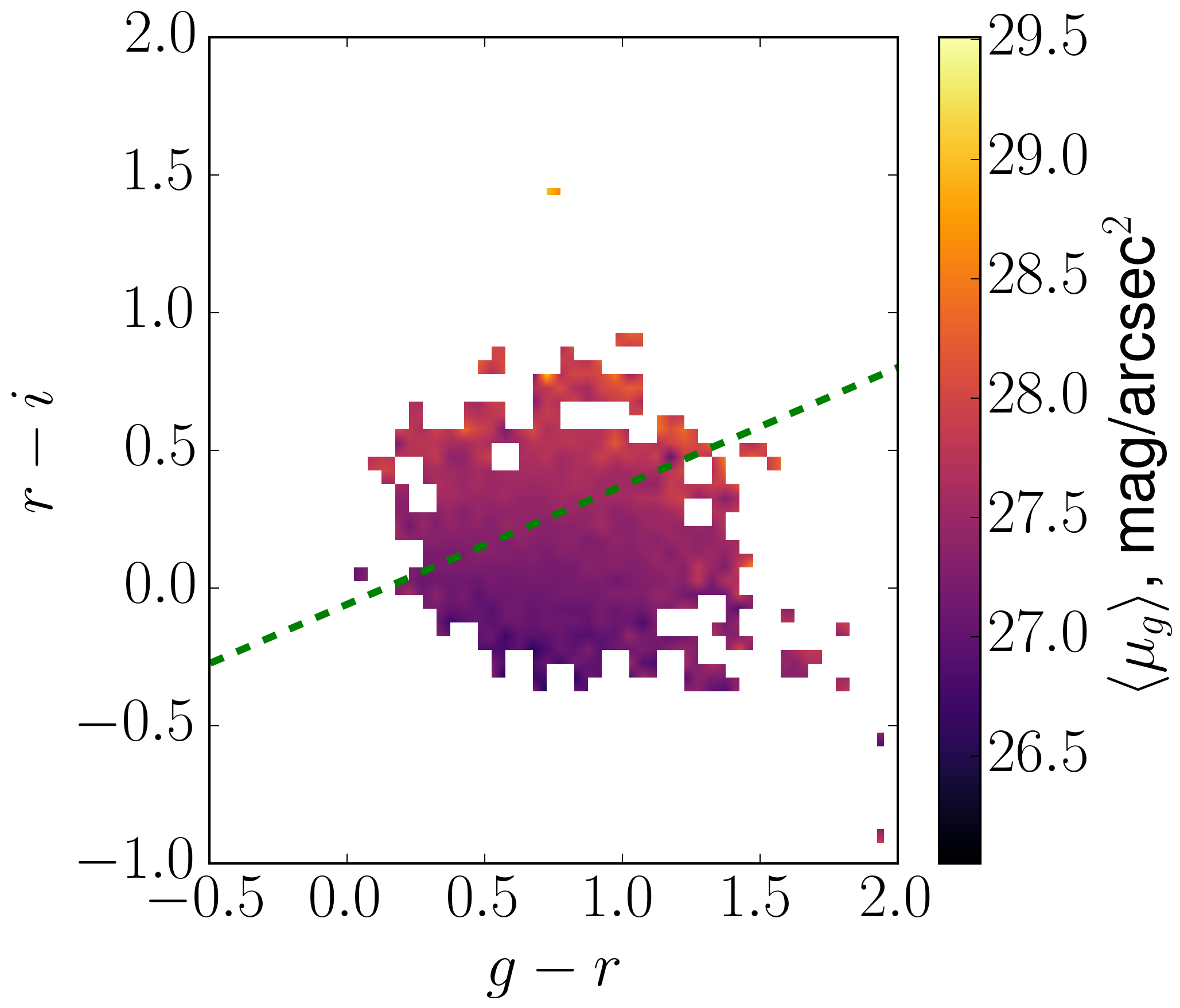}
    \end{minipage} %
    \begin{minipage}{0.45\textwidth}
        \includegraphics[width=1.0\textwidth]{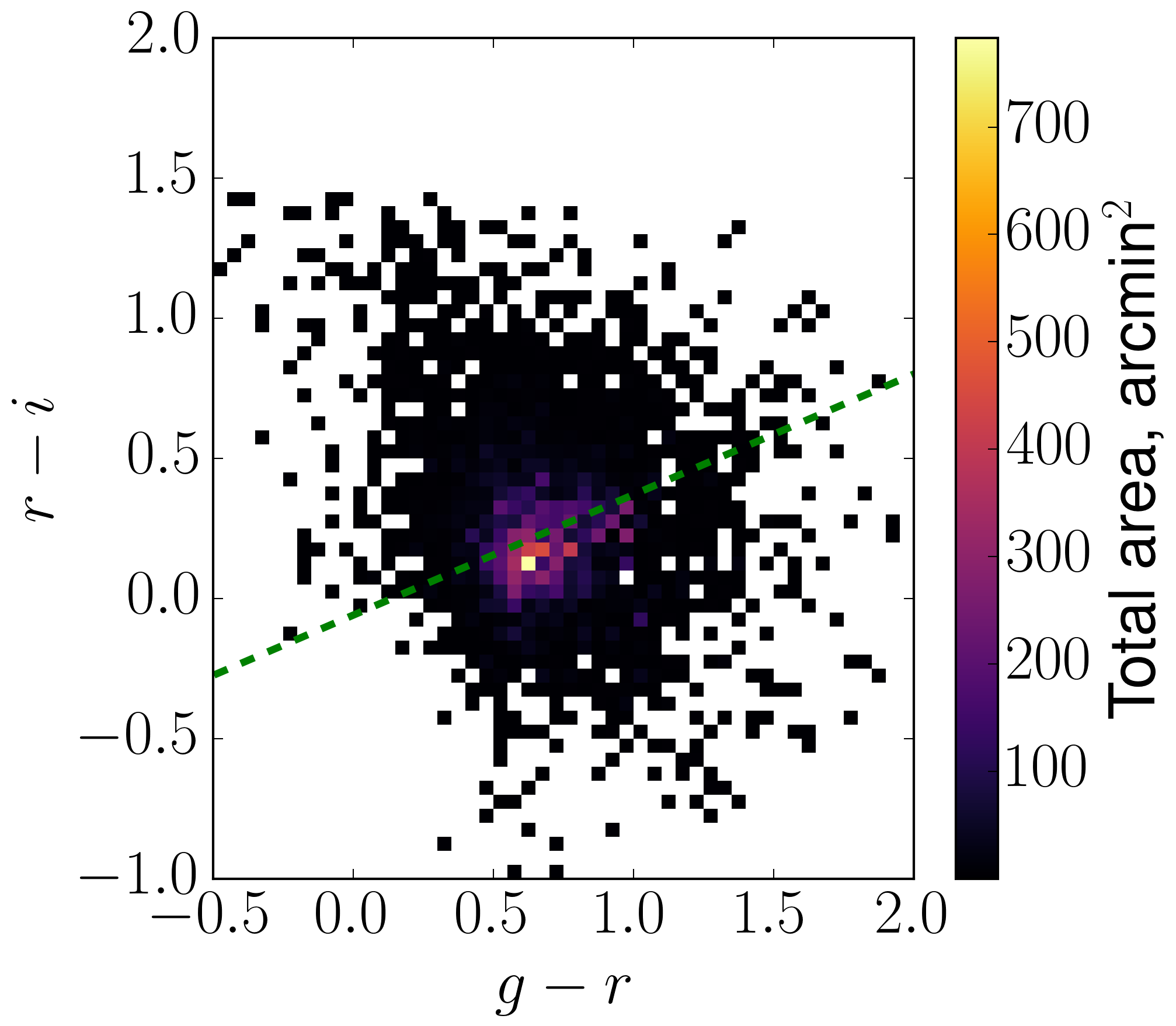}
    \end{minipage}
    \caption{2D distributions of surface brightness (\textit{left}) and area (\textit{right}) depending on the cirrus colours.}
    \label{fig:mu_area_cols}
\end{figure*}

 
An additional argument to support the consistency between our results and those of by~\cite{Roman_etal2020} comes from the analysis of the filament colours depending on the average surface brightness and the area of the filaments. Fig.~\ref{fig:mu_area_cols} shows the corresponding distributions. As can be seen, the larger and brighter the filament, the likelier its colours fall within the limits determined by~\cite{Roman_etal2020}.

\begin{figure*}
    \begin{center}    
    \includegraphics[width = 0.95\textwidth]{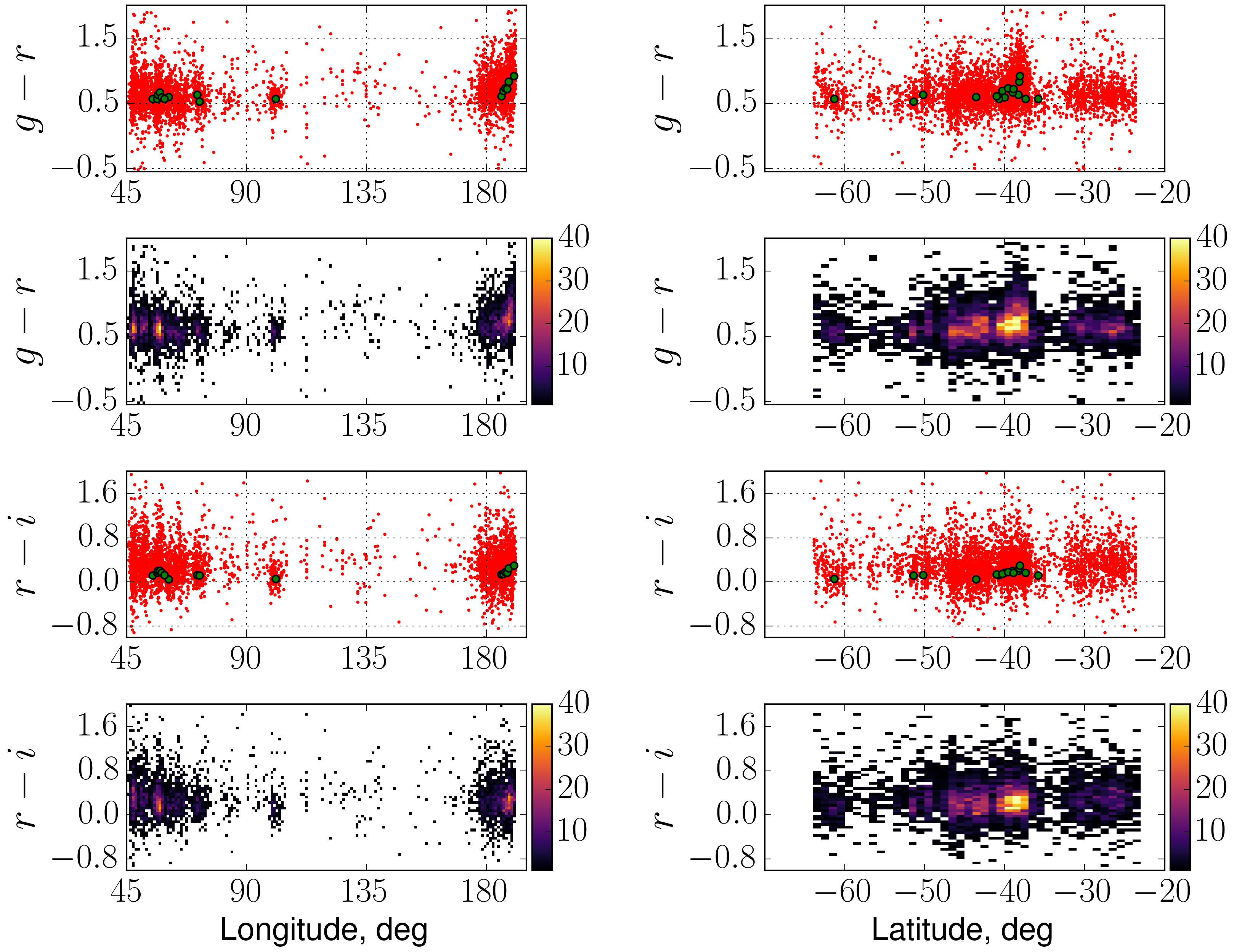}
    \end{center}
    \caption{Distributions of the cirrus  colours $g-r$ (the first two rows) and $r-i$ (the third and fourth rows) colours depending on the galactic longitude (left column) and latitude (right column). First and third rows show usual scatter plots. Second and forth rows show the corresponding 2D histogram by the number of filaments.}
    \label{fig:col_coordinates}
\end{figure*}

\begin{figure*}
    \begin{center}    
    \includegraphics[width = 0.95\textwidth]{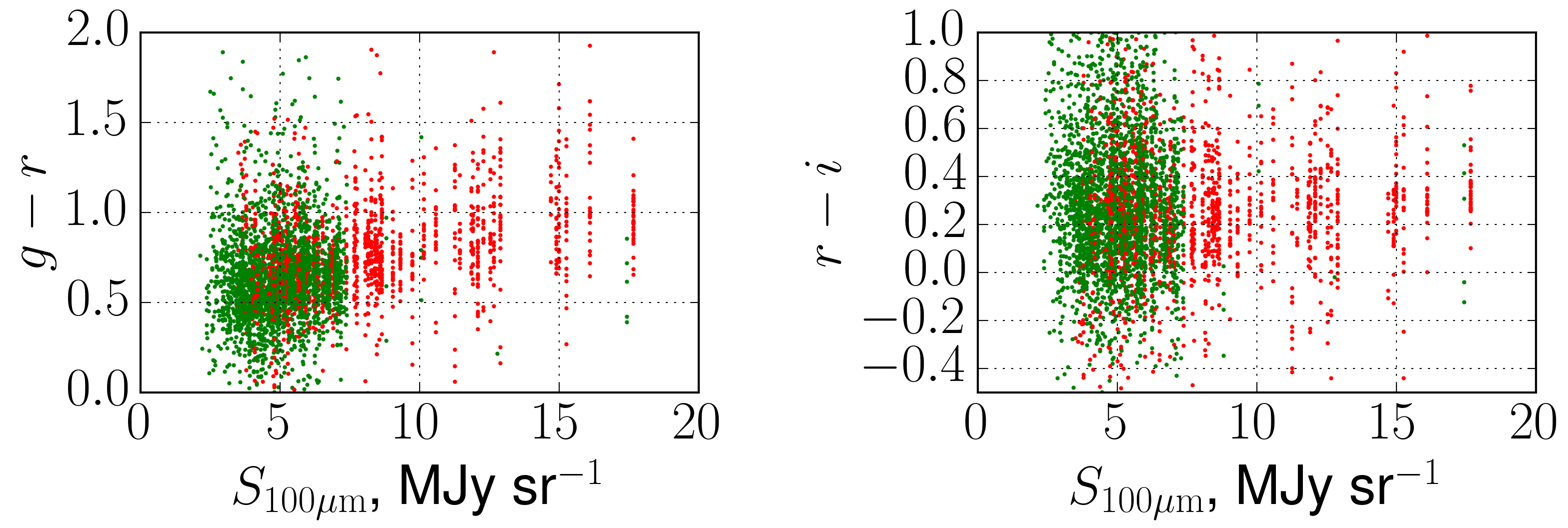}
    \end{center}
\caption{Distributions of the cirrus colours depending on the average far IR emission in the 100 $\mu m$ IRIS band. Green points mark the filaments, which are observed in the region with $l<180$ deg, while red points mark those that have $l \geqslant 180$ deg.}
\label{fig:IRIS}
\end{figure*}

It is also worthwhile to consider the distribution of the cirrus  colours over galactic coordinates. In Fig.~\ref{fig:col_coordinates} we present the distribution of the cirrus colours for all filaments over galactic latitude and longitude (left and right columns, respectively). Each even row shows the distribution as is, while each odd row shows the corresponding 2D histogram by the number of filaments with a bin size of over 1 degree along the $x$-axis and a bin with a colour of 0.05. As can be seen, there is almost no dependence on the coordinates, which is consistent with the results of ~\cite{Roman_etal2020}, thus we confirm the result for individual filaments. One exception is a clear trend at $l\approx180$ deg, where the cirrus clouds become redder. This is connected with an increase of the dust column density in the region, as shown in Fig.~\ref{fig:IRIS}, where we present the distribution of the colours depending on the average far IR emission in the 100 $\mu m$ IRIS band. A similar tendency was found by~\cite{Roman_etal2020} for their clouds, and we confirm their result for individual filaments. The density maps presented in Fig.~\ref{fig:col_coordinates}, also show that the peaks of filaments' distribution appear near the values measured by~\cite{Roman_etal2020}. 

\section{Conclusions}
\label{sec:conclusions}
In the present work, we studied the colour properties of the optical cirrus in Stripe 82 data. The work is inspired by study of~\cite{Roman_etal2020}, where the cirrus colour properties were investigated using the same Stripe 82 data, but only for the largest cirrus clouds. \cite{Roman_etal2020} manually selected some areas of Stripe 82 that contain cirrus clouds and analysed the distribution of the fluxes and colours of all the pixels in those selected areas, then filtered them from all non-cirrus sources of light. Here, we adopted a different approach and tried to identify individual cirrus filaments under the assumption that they can be described as extended objects, the surface brightness of which is greater in each pixel than some specific value determined from the value of background noise (29 mag arcsec$^{-2}$ in the $r$ band for our data). Such a definition of filaments allows one to track the structure of the clouds more accurately, and in particular, measure the colour variance over filaments that constitute the same cloud, for instance. 
\par 
To identify filaments in Stripe 82, we carried out a masking procedure, then selected all sources with $\mu_r$ < 29 mag arcsec$^{-2}$ and visually inspected each of these sources to verify whether or not they appear due to the cirrus scattered light. The latter step is required since the data is contaminated by various sources and extended wings of the PSF. As a result, we marked about 6.4 square degrees of the whole Stripe 82 area as the area dominated by the cirrus scattered light. 
\par 
Since the annotation process of the cirrus is so time consuming, we tested the possibility of optimising it using machine learning methods. We trained a suitable neural network~\textcolor{black}{using the results of manual cirrus annotation as a training sample} and analysed how the training setup (encoder model, training strategy, window size, etc.) affects the results of annotation. We found that models with the MobileNetV2 encoder demonstrate
the highest performance and intersection over union metric value IoU~$=0.576$, which is comparable to the IoU achieved by a human expert (one of the authors). This proves that machine learning methods can be used to solve the problem of cirrus identification. In particular, creating catalogues of cirrus filaments such as those presented by~\cite{2020MNRAS.492.5420S}.
\par 
The resulting sample of identified filaments consists of mostly dim and small features with typical surface brightness  about $\mu_g\approx27$ mag/arcsec$^2$ and area of about $1$ arcmin$^2$.
Since the values differ by an order of a magnitude from those typically considered in previous works, we pay special attention to measuring the optical colours of such features. To this end, we carried out a series of mock simulations, injecting artificial extended sources with a priori known colours into Stripe 82 data. We compared true versus measured  colours for such sources and studied the dependence of the measurement error on the surface brightness and area of the filament. As a result, we identified several pitfalls in the analysis of individual filaments, which should be accounted for in future studies of very faint extended objects (including low surface brightness features around galaxies):
\begin{enumerate}
    \item The linear fitting method for colour estimation does not allow one to retrieve the actual colours of the filaments. Instead, one should use Gaussian fitting suggested by~\cite{Roman_etal2020}.
    \item There is a clear dependence of the colour measurement error on the surface brightness, which is rather expected. However, it is important that the dependence  works in an \textit{average} way, that is, even for bright filaments, some may still have large errors on colours, greater than $0.1-0.2$. \textcolor{black}{At the same time, for most filaments of 26 mag/arcsec$^2$ and brighter, the error of colour measurement is smaller than 0.05. Such bright filaments are most likely to be identified by their true colours in Stripe 82. For dim filaments, the error increases monotonically up to about 0.2 at 28 mag/arcsec$^2$.}
    \item Comparing the colours measured for fields where only noise presents and the actual Stripe 82 data, we found that the colour measurement error should arise mostly from other factors, not due the noise (flatfielding, background subtraction, etc).
\end{enumerate}

\par
As for the optical colours of the filaments distinguished in Stripe 82 data, we found the following. The observed as is, the distribution of the $g-r$ and $r-i$ colours shows a large spread of values arising due to large errors from the contaminating factors, not from the real dispersion of the filaments' colours. At the same time, for most filaments, their colours cluster at some specific values. The comparison of the results of the mock simulations and the data for real filaments indicates that the real colours of the identified filaments should occupy the following ranges: $0.55 \leqslant g-r \leqslant0.73$ and $0.01\leqslant r-i \leqslant0.33$. These ranges are mostly consistent with those previously found in~\cite{Roman_etal2020}. The colours of the filaments also show the tendency to become close to the values measured by~\cite{Roman_etal2020} as surface brightness or filament area increases.
\par 
Overall, the present work provides a useful framework for a future analysis of the upcoming deep optical surveys like Euclid~\citep{Euclid} or the Vera C. Rubin Observatory (former LSST,~\citealt{LSST}). We expect that Galactic cirrus filaments can be identified and studied in these surveys using similar techniques to those developed in this work. 

\section*{Acknowledgements}
We acknowledge financial support from the Russian Science Foundation (grant no. 20-72-10052). Within this grant we performed the following parts of the work: the data analysis (including manual and automatic cirrus segmentation), software development, the creation of the neural network, colour measurements, and the numerical tests.  
JR acknowledges support from the State Research Agency (AEI-MCINN) of the Spanish Ministry of Science and Innovation under the grant "The structure and evolution of galaxies and their central regions" with reference PID2019-105602GB-I00/10.13039/501100011033. JR also acknowledges funding from University of La Laguna through the Margarita Salas Program from the Spanish Ministry of Universities ref. UNI/551/2021-May 26, and under the EU Next Generation. The funding of these grants was used for the data preparation.
\par
We also thank the anonymous referee for the review and appreciate the comments, which allowed us to improve the quality of the publication.
\par 
Funding for the Sloan Digital Sky 
Survey IV has been provided by the 
Alfred P. Sloan Foundation, the U.S. 
Department of Energy Office of 
Science, and the Participating 
Institutions. 
\par
SDSS-IV acknowledges support and 
resources from the Center for High 
Performance Computing  at the 
University of Utah. The SDSS 
website is www.sdss.org.
\par
SDSS-IV is managed by the 
Astrophysical Research Consortium 
for the Participating Institutions 
of the SDSS Collaboration including 
the Brazilian Participation Group, 
the Carnegie Institution for Science, 
Carnegie Mellon University, Center for 
Astrophysics | Harvard \& 
Smithsonian, the Chilean Participation 
Group, the French Participation Group, 
Instituto de Astrof\'isica de 
Canarias, The Johns Hopkins 
University, Kavli Institute for the 
Physics and Mathematics of the 
Universe (IPMU) / University of 
Tokyo, the Korean Participation Group, 
Lawrence Berkeley National Laboratory, 
Leibniz Institut f\"ur Astrophysik 
Potsdam (AIP),  Max-Planck-Institut 
f\"ur Astronomie (MPIA Heidelberg), 
Max-Planck-Institut f\"ur 
Astrophysik (MPA Garching), 
Max-Planck-Institut f\"ur 
Extraterrestrische Physik (MPE), 
National Astronomical Observatories of 
China, New Mexico State University, 
New York University, University of 
Notre Dame, Observat\'ario 
Nacional / MCTI, The Ohio State 
University, Pennsylvania State 
University, Shanghai 
Astronomical Observatory, United 
Kingdom Participation Group, 
Universidad Nacional Aut\'onoma 
de M\'exico, University of Arizona, 
University of Colorado Boulder, 
University of Oxford, University of 
Portsmouth, University of Utah, 
University of Virginia, University 
of Washington, University of 
Wisconsin, Vanderbilt University, 
and Yale University.
\par
This paper has used archival data
from the \textit{Herschel} mission. \textit{Herschel} is an ESA space observatory
with science instruments provided by European-led Principal Investigator consortia and with important participation from NASA.

\section*{Data availability}
The data underlying this article will be shared on reasonable request to the corresponding author. The catalogue of distinguished filaments is available online at ~\url{https://physics.byu.edu/faculty/mosenkov/data}.


\bibliographystyle{mnras}
\bibliography{main}

\appendix
\section{Metrics of cirrus segmentation models}
\label{sec:appendix}
\begin{table*} 
	\caption{Results gathered on the conducted training experiments. It lists the encoder model (MobileNetV2, ResNet50V2, U-Net), training strategy (\guillemotleft training from scratch\guillemotright, \guillemotleft transfer learning\guillemotright, \guillemotleft fine-tuning\guillemotright), window size, input tensor spatial size, number of annotated classes, class weights for background, cirrus and other extended sources if 3 classes are considered, learning rate and IoU, precision, recall for all tests fields for cirrus class. To train all models, we selected 200 square windows from each of the 200 training fields and 100 windows from each of the 50 validation fields.} 
	\centering
    \begin{tabular}{l l c c c c l c c c}
        \hline \hline
        Encoder model  & Training strategy & $w$ (pixel), & $w_{\mathrm{in}}$ (pixel) & $n_{\mathrm{c}}$  & $\overline{\omega_{\mathrm{c}}}$  & $r$ & IoU & precision & recall \\
\hline
MobileNetV2 &training from scratch    &$128$ &$128$   &$2$     &$(1, 1)$&$0.001$  &$0.261$ &$\bmath{0.846}$ &$0.274$  \\
MobileNetV2 &training from scratch    &$128$ &$128$   &$2$     &$(1, 2)$&$0.001$  &$0.437$ &$0.702$ &$0.536$  \\
MobileNetV2 &training from scratch    &$128$ &$128$   &$2$     &$(1, 4)$&$0.001$  &$0.416$ &$0.64$  &$0.543$  \\
\hline
MobileNetV2 &training from scratch    &$224$ &$224$   &$2$     &$(1, 1)$&$0.001$  &$0.44$  &$0.626$ &$0.597$  \\
MobileNetV2 &training from scratch    &$224$ &$224$   &$2$     &$(1, 2)$&$0.001$  &$0.451$ &$0.619$ &$0.624$  \\
MobileNetV2 &training from scratch    &$224$ &$224$   &$2$     &$(1, 4)$&$0.001$  &$0.441$ &$0.623$ &$0.601$  \\
\hline
MobileNetV2 &training from scratch    &$448$ &$448$   &$2$     &$(1, 1)$&$0.001$  &$0.554$ &$0.721$ &$0.706$  \\
MobileNetV2 &training from scratch    &$448$ &$448$   &$2$     &$(1, 2)$&$0.001$  &$\bmath{0.559}$&$0.678$ &$0.761$  \\
MobileNetV2 &training from scratch    &$448$ &$448$   &$2$     &$(1, 4)$&$0.001$  &$0.48$  &$0.552$ &$\bmath{0.788}$  \\
\hline
MobileNetV2 &training from scratch    &$4\times 224$  &$224$   &$2$     &$(1, 1)$&$0.001$  &$0.482$ &$0.774$ &$0.561$  \\
MobileNetV2 &training from scratch    &$4\times 224$  &$224$   &$2$     &$(1, 2)$&$0.001$  &$0.513$ &$0.674$ &$0.683$  \\
MobileNetV2 &training from scratch    &$4\times 224$  &$224$   &$2$     &$(1, 4)$&$0.001$  &$0.498$ &$0.806$ &$0.566$  \\
\hline
\hline
MobileNetV2 &training from scratch    &$448$ &$448$   &$3$     &$(1, 1, 1)$ &$0.001$  &$0.538$ &$0.617$ &$\bmath{0.807}$  \\
MobileNetV2 &training from scratch    &$448$ &$448$   &$3$     &$(1, 2, 1)$ &$0.001$  &$\bmath{0.559}$&$\bmath{0.649}$ &$0.802$  \\
MobileNetV2 &training from scratch    &$448$ &$448$   &$3$     &$(1, 4, 1)$ &$0.001$  &$0.543$ &$0.637$ &$0.786$  \\
\hline
\hline
MobileNetV2 &fine-tuning &$448$ &$448$   &$2$     &$(1, 1)$&$0.001$  &$0.512$ &$0.682$ &$0.673$  \\
MobileNetV2 &fine-tuning &$448$ &$448$   &$2$     &$(1, 1)$&$0.0005$  &$0.551$ &$0.668$ &$0.758$  \\
MobileNetV2 &fine-tuning &$448$ &$448$   &$2$     &$(1, 1)$&$0.00025$&$0.47$  &$\bmath{0.875}$ &$0.504$  \\
MobileNetV2 &fine-tuning &$448$ &$448$   &$2$     &$(1, 2)$&$0.001$  &$0.548$ &$0.739$ &$0.68$   \\
MobileNetV2 &fine-tuning &$448$ &$448$   &$2$     &$(1, 2)$&$0.0005$  &$\bmath{0.576}$&$0.676$ &$\bmath{0.796}$  \\
MobileNetV2 &fine-tuning &$448$ &$448$   &$2$     &$(1, 2)$&$0.00025$&$0.463$ &$0.762$ &$0.542$  \\
MobileNetV2 &fine-tuning &$448$ &$448$   &$2$     &$(1, 4)$&$0.001$  &$0.541$ &$0.694$ &$0.711$  \\
MobileNetV2 &fine-tuning &$448$ &$448$   &$2$     &$(1, 4)$&$0.0005$  &$0.573$ &$0.701$ &$0.758$  \\
MobileNetV2 &fine-tuning &$448$ &$448$   &$2$     &$(1, 4)$&$0.00025$&$0.458$ &$0.606$ &$0.653$  \\
\hline
MobileNetV2 &transfer learning        &$448$ &$448$   &$2$     &$(1, 1)$&$0.001$  &$0.414$ &$0.621$ &$0.554$  \\
MobileNetV2 &transfer learning        &$448$ &$448$   &$2$     &$(1, 1)$&$0.0005$  &$0.397$ &$0.604$ &$0.537$  \\
MobileNetV2 &transfer learning        &$448$ &$448$   &$2$     &$(1, 1)$&$0.00025$&$0.415$ &$0.622$ &$0.554$  \\
\hline
ResNet50V2  &training from scratch    &$448$ &$448$   &$2$     &$(1, 1)$&$0.001$  &$0.55$  &$0.683$ &$0.738$  \\
ResNet50V2  &training from scratch    &$448$ &$448$   &$2$     &$(1, 2)$&$0.001$  &$0.57$  &$0.749$ &$0.706$  \\
ResNet50V2  &training from scratch    &$448$ &$448$   &$2$     &$(1, 4)$&$0.001$  &$0.54$  &$0.66$  &$0.748$  \\
\hline
U-Net   &training from scratch    &$448$ &$448$   &$2$     &$(1, 4)$&$0.001$  &$0.155$ &$0.773$ &$0.163$  \\
U-Net   &training from scratch    &$448$ &$448$   &$2$     &$(1, 8)$&$0.001$  &$0.415$ &$0.684$ &$0.514$  \\
U-Net   &training from scratch    &$448$ &$448$   &$2$     &$(1, 16)$   &$0.001$  &$0.388$ &$0.474$ &$0.681$  \\

    \end{tabular} 
    \label{tab:nn_metrics}
\end{table*}   

\end{document}